# Banking system stability: A global analysis of cybercrime laws

**Douglas Cumming**
School of Business
Stevens Institute of Technology, 525 River St., Hoboken, NJ 07030, United States
Email: dcumming@stevens.edu

**My Nguyen**
School of Economics, Finance and Marketing
RMIT University, 445 Swanston Street, Melbourne, VIC 3000, Australia
Email: my.nguyen@rmit.edu.au

**Anh Viet Pham**
School of Economics, Finance and Marketing
RMIT University, 445 Swanston Street, Melbourne, VIC 3000, Australia
Email: anh.pham2@rmit.edu.au

**Ama Samarasinghe**
School of Accounting, Information Systems and Supply Chain
RMIT University, 445 Swanston Street, Melbourne, VIC 3000, Australia
Email: ama.samarasinghe@rmit.edu.au

**Abstract**

We examine the role of cybercrime legislation around the world in shaping the stability of the banking system. We compile a novel dataset covering the enactment of cybercrime legislation in 132 developed and developing countries to empirically test this research question. We find that the enactment of cybercrime laws enhances the stability of the banking sector. This key finding holds across a comprehensive suite of robustness tests, including alternative measures of bank stability and model specifications. We document significant cross-sectional heterogeneity, with the effect being more pronounced in countries with heavier penalties for illegal cyber activities and legal frameworks that hold banks accountable for their cybersecurity practices. In addition, the positive impact is stronger in jurisdictions with greater international legal cooperation and effective enforcement mechanisms. We further investigate two channels (i.e., funding liquidity and operational risk) through which cybercrime laws may influence bank stability. Our results indicate that these laws can significantly bolster bank stability by enhancing funding liquidity and mitigating operational risk. Overall, our study highlights the crucial role of cybercrime legislation in fostering a secure and resilient banking environment. It offers new insights into how these laws contribute to bank stability on both individual and systemic levels.

**Acknowledgements**

We are grateful for the comments and guidance from Lemma Senbet (the area editor), Rosalie Tung (the editor-in chief), Ruth Aguilera (the deputy editor), and two anonymous reviewers. The paper has also benefited from comments and discussions with Kristle Romero Cortes, Huu Nhan Duong, Harrison Hong, Thanh Huynh, Phong Ngo, Jason Tian, and Sean Wu, as well as conference and seminar participants at the 2023 New Zealand Finance Conference, the 2023 FIRN-ANU Banking and Financial Stability Meeting, and the 2024 Academy of International Business Annual Conference. We would like to thank Giang Nguyen, Hoa Phan, Stamatios Tsigos, and Lynn Zhou for their excellent research assistance. All errors are ours.

# Banking system stability: A global analysis of cybercrime laws

**Abstract**

We examine the role of cybercrime legislation around the world in shaping the stability of the banking system. We compile a novel dataset covering the enactment of cybercrime legislation in 132 developed and developing countries to empirically test this research question. We find that the enactment of cybercrime laws enhances the stability of the banking sector. This key finding holds across a comprehensive suite of robustness tests, including alternative measures of bank stability and model specifications. We document significant cross-sectional heterogeneity, with the effect being more pronounced in countries with heavier penalties for illegal cyber activities and legal frameworks that hold banks accountable for their cybersecurity practices. In addition, the positive impact is stronger in jurisdictions with greater international legal cooperation and effective enforcement mechanisms. We further investigate two channels (i.e., funding liquidity and operational risk) through which cybercrime laws may influence bank stability. Our results indicate that these laws can significantly bolster bank stability by enhancing funding liquidity and mitigating operational risk. Overall, our study highlights the crucial role of cybercrime legislation in fostering a secure and resilient banking environment. It offers new insights into how these laws contribute to bank stability on both individual and systemic levels.

*"This year (2020), while virtually all sectors of the global economy fell victim to cybercrime of one kind or another, no sector was more regularly targeted than the financial sector…. Every organization– providers of financial services, in particular – must remain vigilant in the face of these evolving threats. It is critical that organizations maintain a continuous dialogue with law enforcement to ensure a rapid response in the event of an incident."*

***Jonah Force Hill***

***Executive Director of the U.S. Secret Service Cyber Investigations Advisory Board***

# INTRODUCTION

Cyberattacks on information and communication technology systems are increasing worldwide, with the financial sector being a primary target. These attacks have caused major financial losses and operational disruptions. In 2012, a series of coordinated distributed denial of service (DDoS) attacks against six major American banks, including Bank of America, Citigroup, JP Morgan Chase, PNC Bank, U.S. Bank, and Wells Fargo, crippled online services, leaving millions of customers unable to access accounts or pay bills, leading to widespread frustration. The attacks have a significant impact on the affected banks in terms of mitigation expenses, potential lost business, and reputation. In another incident in 2024, an anonymous hacker group breached 20 Iranian banks, gaining unauthorized access to millions of personal accounts and credit cards; the targeted banks reportedly paid millions in ransom to prevent the hackers from releasing data, including personal account information. In light of such attacks, governments worldwide have enacted cybercrime laws to mitigate cyber risk and minimize cyber threats.

According to the United Nations Office on Drugs and Crime (UNODC) (2013), cybercrime laws provide a critical legal framework to prosecute illegal cyber activities. Through punitive measures, they discourage the intent to commit cyber offenses, which, in turn, reduces the frequency and severity of cyber incidents. Cybercrime laws also encourage banks to invest in robust cybersecurity measures and adopt a proactive stance against potential cyber threats. Thus, cybercrime laws are expected to play a pivotal role in minimizing cyber risk, thereby enhancing banking system stability. Nonetheless, moral hazards can manifest if banks become overly reliant on the perceived security provided by cybercrime laws. They may allocate fewer resources towards their cybersecurity layers, which, in turn, exposes them to a higher risk of cyber incidents, negatively impacting their stability. Collectively, these contrasting arguments motivate an important research question as to what extent the enactment of cybercrime laws affects banking system stability.

Our paper addresses this question by employing a unique dataset on the enactment of cybercrime laws in 132 developed and developing countries. The enactment of cybercrime laws, being exogenous to individual banks and determined by each country's legislative bodies, provides a valid setting for our analysis. Furthermore, given the staggered enactment of cybercrime laws worldwide, we can control any global trends that may influence banking activities. Our study reveals a significant positive impact of cybercrime laws on bank stability. In terms of economic significance, on average, individual bank stability increases by 0.69%, and banking system stability increases by 25.58% following the enactment of cybercrime laws. To address potential endogeneity concerns, we use a stacked event-by-event approach. These results, consistent with our baseline findings, suggest the causal impact of cybercrime laws on bank stability.

We further investigate the heterogeneity of cybercrime laws on banking system stability across different regulatory attributes. First, we find that the positive impact of cybercrime laws on bank stability is stronger among jurisdictions imposing heavier penalties on cyber offenses. Second, cybercrime laws that explicitly hold banks accountable for their cybersecurity practices have a more pronounced positive impact on stability. Third, the positive relationship between cybercrime laws and bank stability is strengthened among countries that engage in more bilateral extradition treaties with others. Finally, we find that the favorable effect of cybercrime laws on bank stability is more pronounced in countries where laws are applied and enforced equally.

To gain insight into how cybercrime laws affect bank stability, we investigate two channels: funding liquidity and operational risk. Effective cybercrime laws are likely to play a vital role in enhancing depositors' and investors' confidence in the banking system, leading to a surge in demand for bank funding. Under cybercrime laws, banks may need to establish robust cybersecurity protocols, provide employee training, and regularly update security systems. This approach helps mitigate the risk of cyber incidents and potential damage, which essentially reduces operational risk and thus bolsters overall bank stability.

Nonetheless, in light of the lower risk of cyber incidents in the presence of cybercrime laws, banks may neglect strong cyber defenses; if so, a cyber breach occurs, its aftermath will likely be more detrimental due to increased risk of operational disruptions. Implementing rigorous cybersecurity measures may create operational challenges and, hence, operational risks for banks. Collectively, these arguments suggest that cybercrime laws can cause a reduction in funding liquidity while elevating operational risks for banks. We utilize path analysis to investigate this and document a significant positive relationship between the enactment of cybercrime laws and funding liquidity, as well as a reduction in bank operational risk. Overall, we find that cybercrime laws can strengthen the banking system's stability through funding liquidity and operational risk reduction.

Our study offers several contributions to literature. We add to the literature on the role of cybercrime laws in banking activities worldwide. The extant literature focuses on the effect of cyber risk on firm stock return (Florackis, et al., 2022), idiosyncratic volatility (Nazil et al., 2021), supply chain (Crosignani, et al., 2023), firm reputation (Kamiya, et al., 2021), payment systems (Eisenbach et al., 2022), and individual bank risk (Duffie and Younger, 2019). This paper is the first to examine whether the enactment of cybercrime laws can enhance financial stability at both bank- and systemic-levels. We also investigate the channels through which these laws can affect bank stability, adding to the expanding body of research on macro-determinants of bank funding liquidity and operational risk (Berger et al., 2023; Raz et al., 2022).

Second, this study enhances the literature investigating the factors influencing financial stability. Prior research has emphasized bank-specific characteristics and macro-level variables, such as GDP growth, as main factors affecting banking system stability (Goetz, 2018; Shim, 2019). At the systemic level, the literature revolves around conceptual and theoretical frameworks that define systemic risk and the design of prudential regulatory tools in preventing costly banking system crises (Acharya, 2009; Adrian and Brunnermeier, 2016). There is limited literature concerning the impact on bank stability stemming from non-traditional sources, such as cyber risk (Thakor, 2023). This paper extends the current literature by highlighting the role of cybercrime laws in mitigating potential cyber breaches facing the financial services sector.

Third, this study contributes to the literature on multinational banks (MNBs). Prior studies have explored the role of institutional and market differences, governance structures, and regulatory frameworks in shaping MNB lending, liquidity creation, and strategies (Berger et al., 2024; Chen et al., 2022). Amid escalated cyber threats, MNBs face unique risks due to their extensive cross-border operations, reliance on global financial networks, and exposure to contagion, where a cyberattack on one entity can spread across subsidiaries and counterparties, disrupting trade, payments, and capital flows. These risks are heightened by cross-country differences in cybercrime laws, breach severity, and banking stability (Adelmann et al., 2020; Madan et al., 2023). Thus, understanding the impact of cybercrime laws on banking stability is highly relevant from an international business perspective, as it sheds light on regulatory mechanisms that influence both MNB resilience and systemic financial stability.

The remainder of the paper is structured as follows. The next section reviews the related literature and develops hypotheses. Thereafter, we provide details on data collection, variable descriptions, and a presentation of the main results and additional analyses. The final section offers concluding remarks.

## LITERATURE REVIEW AND HYPOTHESIS DEVELOPMENT

In an increasingly digitalized financial services industry, cyberattacks are a distinct source of risk that challenges banking system stability. Unlike traditional risks, they involve malicious intent targeting the confidentiality, availability, and integrity of financial systems, setting them apart from operational risks that occur randomly or seasonally, such as natural disasters or technical failures (Eisenbach et al., 2022). Technology-driven linkages facilitate their spread across institutions and thereby exacerbate systemic risk, whereas financial risks typically stem from solvency or liquidity shocks arising from business interactions. Moreover, uncertainty around the detection and attribution of cyber events amplifies their disruptive potential, further distinguishing them from other financial risks (Adelmann et al., 2020; Duffie and Younger, 2019).

Central to this study's theoretical framework is the institutional theory developed by North (1990) and Scott (1995). North (1990) defines institutions as "the humanly devised constraints that structure political, economic, and social interaction," encompassing formal constraints (e.g., laws, regulations) and informal constraints (e.g., norms, conventions) that collectively shape incentive structures within societies and economies. Scott (1995) categorizes these institutions into three pillars: the regulative pillar (formal rules and laws), the normative pillar (norms and values), and the cultural-cognitive pillar (shared beliefs and common understandings). The implementation of cybercrime laws represents the regulative pillar, providing a legal framework to address and combat cyber-attacks. As a significant deterrent, these laws discourage cybercriminal activity through the likelihood of prosecution and punishment, thereby reducing cyber incidents, and ultimately enhancing financial stability. This perspective aligns with broader literature emphasizing the importance of legal deterrents in reducing cybercrime and its associated risks (Barton et al., 2013; Levi, 2017).

In addition, the significance of cybercrime laws goes beyond punishing cybercriminals. They hold both cybercriminals and banks accountable for cyber incidents and cybersecurity practices. For example, while criminalizing illegal cyber activities (e.g., unauthorized access, hacking), the Cybersecurity Law of the People's Republic of China (2017) also requires companies to implement substantial cybersecurity measures; failure to comply results in monetary penalties and even business suspensions. Such requirements encourage banks to invest in robust cybersecurity, reducing exposure to successful cyber-attacks and supporting banking system stability.

However, stringent regulations may create moral hazard, whereby increased oversight and managerial myopic actions inadvertently incentivize complacency in risk management (Chen et al., 2024; John et al., 2000). This can similarly occur with cybercrime laws. When banks perceive that these laws mitigate their cyber risks, for example, they may be less vigilant in practicing sound cyber risk management, raising the likelihood of breaches and losses and threatening the stability of individual banks and the broader financial system.

While cybercrime laws aim to reduce the frequency and severity of cyberattacks on financial institutions, they also impose compliance and cybersecurity costs (Jia et al., 2018). Banks are expected to establish cybersecurity programs which involve hiring experts, conducting routine evaluations, and maintaining comprehensive documentation. These costs can be significant, especially for smaller banks, and may undermine financial stability. Considering these competing arguments, we propose the following null hypothesis:

*H1: There is no significant relationship between the enactment of cybercrime laws and banking system stability.*

The distinct characteristics of cyber risk carry significant implications for financial stability, especially for funding liquidity and contagion. Cyberattacks can immobilize capital and liquidity, hindering banks' ability to meet obligations and perform core activities due to compromised data or systems. This can disrupt payments or restrict account access (Eisenbach et al., 2022). Such disruption may trigger panic-driven withdrawals even from unaffected institutions, intensifying liquidity pressures and undermining market confidence (Altermatt et al., 2024; Caballero and Simsek, 2013; Kashyap and Wetherilt, 2019).

Since cyber incidents can trigger funding liquidity crunches, lower cyber risk following the adoption of cybercrime laws should reduce the likelihood of runs. Customers perceiving a lower risk of losses from hacking are more comfortable making deposits, strengthening funding liquidity (Duffie and Younger, 2019; Eisenbach et al., 2022). These perceptions may also improve a bank's access to wholesale funding and interbank lending.

However, moral hazard could arise if banks become overly reliant on regulatory protections and weaken their internal cybersecurity practices. In such an environment, stakeholders may hesitate to provide funding (e.g., deposits, lending), wary of banks' reduced vigilance against cyber threats. Based on these two-sided arguments, we conjecture the following null hypothesis:

*H2: The enactment of cybercrime laws worldwide does not impact bank stability through the funding liquidity channel.*

Operational risk - encompassing losses due to internal processes, personnel, or external events - poses a significant challenge to the stability of financial institutions. Cyberattacks emerge as a substantial source of operational risk, compromising sensitive customer information, operational disruptions, and financial losses (Jamilov et al., 2023). Recent advancements in financial service offerings, such as extended operating hours and compressed clearing and settlement windows, have inadvertently increased susceptibility to cyber shocks. A distinctive aspect of cyberattacks is their potential for maximum disruption, shaped by attackers' insights into payment system architecture, targeted institutions, and interconnections within the payment network. Traditional risk management frameworks focus on counterparty relationships and interbank borrowing (Allen and Gale, 2000), yet cyber vulnerabilities can propagate through unanticipated network pathways not captured by conventional assessments. Correlated impairments from cyber-induced disruptions can further exacerbate systemic vulnerabilities, underscoring the interconnected nature of cyber risk within financial networks (Erol and Vohra, 2022).

Cybercrime laws provide a proactive means to mitigate operational risks from cyber-attacks and promote banking sector stability. By establishing a legal framework to prosecute illegal cyber activity, these laws deter crime, reducing the frequency and severity of attacks. Regulations may also require banks to implement robust cybersecurity protocols, strengthening defences against data breaches and online fraud, further limiting operational risk.

However, perceived protection from government policies can create moral hazard (Dam and Koetter, 2012), prompting some banks to reduce investments in security systems and training. Compliance also imposes direct costs (e.g., cybersecurity technology, legal fees, ongoing audits) and indirect costs (e.g., reallocating human capital to compliance tasks); if poorly managed, these adjustments may inadvertently increase operational risk. Collectively, these contrasting arguments give rise to the following null hypothesis:

*H3: The enactment of cybercrime laws does not impact bank stability through operational risk.*

## DATA COLLECTION AND VARIABLE DESCRIPTIONS

**Data, sampling procedure, and sample selection**

We obtain commercial banks' accounting data from Fitch Connect (Fitch Solutions), cybercrime legislation from the United Nations Conference on Trade and Development's Cybercrime Legislation Worldwide database, market-based bank stability from Datastream, and macroeconomic status and national governance indices from the World Bank's World Development Indicators and Kaufmann et al. (2019), respectively.

To build the final dataset, we (i) exclude banks with fewer than three consecutive annual observations and those with negative assets, loans, or deposits, and (ii) treat target and acquiring banks separately when reported distinctly, excluding targets if unconsolidated data are unavailable after a merger with a non-bank acquirer. To prevent survivorship bias, we utilise an unbalanced bank-specific panel covering as many banks as possible, including those not operating throughout the entire sample period. The dataset spans 1990–2021 and comprises 20,050 listed and non-listed commercial banks across 132 developed and developing countries. To construct systemic stability measures, we manually merge the sample of listed commercial banks with the Datastream database using bank names. We exclude non-listed commercial banks from our systemic stability analysis, as Datastream does not provide data for such banks. Therefore, the sample used in our systemic stability analysis contains 645 listed commercial banks in 79 countries.

**The enactment of cybercrime laws**

The United Nations Conference on Trade and Development (UNCTAD) maintains the Cybercrime Legislation Worldwide database, which provides a global overview of national cybercrime laws, including definitions, offences, penalties, enforcement mechanisms, and international cooperation frameworks. It also documents training and awareness initiatives, making it a key resource for analyzing global efforts to combat cybercrime.

As of December 2021, the database covers a comprehensive list of 175 countries. We manually gather data on the year in which each country enforces cybercrime legislation. As of December 2021, 156 countries (80%) have successfully enacted cybercrime laws. Of the remaining countries, 5% have enforced draft legislation, 13% have yet to enact any legislation, and 1% have no data available regarding their enactment status. Noticeably, the pattern of cybercrime law enactment is not uniform across regions. Europe boasts the highest adoption rate of 91%, whereas Africa has the lowest adoption rate of 72%. Of particular interest, nineteen nations had already established cybercrime laws prior to 2000, with Costa Rica leading the world by instituting Section 12 of its Cyber Crime Code as early as 1970. Between 2000 and 2010, an additional 53 countries introduced their own legislation to combat cybercrime. Figure 1 visually represents the global status of cybercrime law enactment as of December 2021.

[Insert Figure 1 here]

**Banking system stability measures**

To examine bank stability, we utilize various measures including the Z-index, the marginal expected shortfall (MES), and the SRISK.

*Z-index*

The Z-index, denoted as *Z_SCORE*, measures a bank's resilience to financial stress. According to Khan et al. (2017), it indicates the number of standard deviations below the mean that a bank's profits would need to decrease to deplete its equity capital. The Z-index is derived as the ratio of the sum of the return on assets and the capital-to-asset ratio to the standard deviation of asset returns. The computation of the Z-index is outlined as follows:

$$Z-index_{i,t} = \frac{ROA_{i,t} + \left(\frac{Equity_{i,t}}{Assets_{i,t}}\right)}{STDEV(ROA_{i,t})} \tag{1}$$

where the return on assets (ROA) is the bank profitability ratio, computed as net income divided by total assets. The standard deviation of asset returns $STDEV(ROA)$ is determined through a three-year rolling window. The Z-index is calculated based on the proportion of ROA and the capital-to-asset ratio to the standard deviation of asset returns. A higher Z-index suggests enhanced stability for a bank, as it indicates a greater likelihood of remaining solvent during episodes of financial stress.

*Marginal expected shortfall*

The marginal expected shortfall (*MES*) for a bank is characterized as the anticipated equity loss in the immediate term, assuming a market decline surpassing its Value-at-Risk (VaR) at the α% confidence level (Acharya et al., 2017). Simply put, the MES measures the additional losses a bank is predicted to endure beyond its VaR in the occurrence of a substantial market downturn. The formula for MES is as follows:

$$MES_{i,t} = E_t(R_{i,t+1} | R_{m,t+1} < q_{\alpha,t}(R_{t+1}) = C) \tag{2}$$

In our investigation, *C* is a representation of the market's "tail risk," assessed through the VaR at the 5th percentile. The day-to-day stock return for bank *i* on day *t* is expressed as $R_{i,t}$, while $R_{m,t}$ stands for the daily market index return. The anticipated shortfall of the index (*ES*) is computed as the projected loss under the condition that the loss surpasses the *C* level. The formula for *ES* is as follows:

$$ES_t = E_t(R_{m,t+1}|R_{m,t+1} < C) \qquad (3)$$

When a bank is part of the market, a bank's marginal expected shortfall is the derivative of the market's expected shortfall with respect to the bank's capitalization or market shares. This shows that *MES* is an indicator of the bank's contribution to systemic risk. *MES* offers various advantages over alternative measures like Value at Risk, including its capacity to consider extreme events without excluding them, its lack of reliance on a normal distribution assumption, and its efficacy in forecasting the performance of the weakest banks during crises, as evidenced in studies such as Acharya et al. (2012). A heightened MES represents diminished bank stability.

*SRISK*

Following Brownlees and Engle (2017), we construct a measure of the expected capital deficit of a financial entity in the event of a prolonged market downturn (*SRISK*). This metric outperforms systemic expected shortfall, as suggested by Acharya et al. (2017), in terms of its predictive accuracy and is free of any structural assumptions. In mathematical terms, *SRISK* is formulated as:

$$SRISK_{i,t} = kD_{i,t} - (1-k)W_{i,t}(1 + LRMES_{i,t}) \qquad (4)$$

In Equation (4), *SRISK* is a function of a bank's size, leverage level, and projected equity loss given a market downturn, denoted as the long-run marginal expected shortfall (*LRMES*). *LRMES* is characterized as the anticipation of a bank's equity multi-period arithmetic return under the condition of a systemic event, expressed as:

$$LRMES_{it} = -E_t(R_{it+1:t+h}|R_{m\ t+1:t+h} < C) \qquad (5)$$

where $R_{it+1:t+h}$ represents the multiperiod arithmetic return of a bank's equity between periods *t+1* and *t+h*. We collect daily data on the book value of debt (*D*) and the market value of equity (*W*) of financial entities within each country, along with their prudential capital fraction (*k*). Using year-end values for each country, we compute the expected capital shortfall for the entire banking system using the *SRISK* measure. To accommodate variations in the size of economies, we normalize this systemic risk measure by the real Gross Domestic Product (GDP) of each country. Table IA1 provides the list of 132 countries and the mean value of bank stability measures in our sample countries.

**Descriptive statistics and correlation matrix**

Table 1 presents descriptive statistics (Panel A) and a correlation matrix (Panel B) for our key variables. The mean natural logarithm of *Z_SCORE* is 3.579, indicating that, on average, profits would need to decline about 35.79 times their standard deviation to fully exhaust bank equity. The mean values for *MES*, calculated at the 5$^{th}$ percentile of value at risk (i.e., *MES_VAR5%*) and *SRISK*, are 0.125 and 0.072, respectively. The average *CYBERLAW* value is 0.811, suggesting that most of the sampled countries enacted cybercrime laws during the study period. Customer deposits, on average, constitute 85.3% of total bank funding. The common equity ratio has a mean value of 11.5%. The average natural logarithm of bank assets is 19.72. Non-interest income represents, on average, 22.4% of the total operating income for the sample, with the 75$^{th}$ percentile at 28.2%.

Table 1, Panel B displays the correlation matrix of variables used in our baseline regression model (Equation (6)). *Z_SCORE* is negatively correlated with the two bank systemic risk measures, including *MES_VAR5%* and *SRISK*. In addition, the enactment of cybercrime law (*CYBERLAW*) shows a positive correlation with *Z_SCORE* and a negative correlation with *MES_VAR5%* and *SRISK*. This finding offers preliminary evidence of a positive association between the enactment of cybercrime law and bank stability.

[Insert Table 1 here]

**CYBERCRIME LAWS AND BANKING SYSTEM STABILITY**

**Baseline results**

As a starting point, we estimate the impact of the enactment of cybercrime laws on banking system stability by employing the following regression model:

$$Bank\ Stability_{i,c,t+1} = \alpha + \beta_1 CYBERLAW_{c,t} + \beta_2 Bank\ Controls_{i,c,t} + \beta_3 Macro\ Controls_{ct} + v_c + \mu_t + \varepsilon_{i,c,t+1} \qquad (6)$$

In Equation (6), $i$, $c$, and $t$ denote bank, country, and year, respectively. The binary variable *CYBERLAW* takes the value 1 if a country has enacted cybercrime laws and 0 otherwise. Bank stability measures include the Z-index (*Z_SCORE*), the marginal expected shortfall estimated at the 5$^{th}$ percentile of value of risk (*MES_VAR5%*), and the systematic risk (*SRISK*). Intuitively, it may take time for cybercrime laws to exert an effect. Hence, we lead all bank stability measures to one year. Bank control variables consist of bank size (*BANK_SIZE*), book equity ratio (*EQUITY_RATIO*), customer deposits ratio (*DEPOSIT_RATIO*), cost efficiency (*COST_EFFICIENCY*), and revenue diversification (*REV_DIV*). Macroeconomic control variables include explicit deposit insurance *(EXP_DEP_INSUR)*, the rule of law (*RULE_OF_LAW*), GDP (*LOG_GDP*), log GDP per capita (*LOG_GDPPERCAP*), and the private credit-to-GDP ratio (*CREDIT_TO_GDP*), accounting for variations in economic development and institutional quality among countries (Barth et al., 2004; Cull et al., 2005; Cope et al., 2012). In the subsequent robustness tests, we will also present results after controlling for additional bank activity restrictions, as well as institutional and other macroeconomic indicators.

The literature identifies bank size as a key determinant of bank stability, though the direction is mixed (Micco et al., 2007). We control for the book-to-equity ratio, as higher capital buffers are associated with greater stability (John et al., 2004). We also include a bank's dependence on customer deposits, measured as total customer deposits to total funding, reflecting capital structure and financial health. Because inefficient management can cause distress, we control for cost efficiency; higher funding costs may signal inefficiency (Berger and DeYoung, 1997). Finally, revenue diversification, capturing the bank's business model, can also affect stability (Dietrich and Wanzenried, 2011). A detailed definition of these variables is provided in Table A1. To control for unobserved heterogeneity, we employ country and year fixed effects ($v_c$ and $\mu_t$, respectively).[1] The error term is captured by $\varepsilon_{i,c,t}$. We winsorize all continuous variables at the 1$^{st}$ and 99$^{th}$ percentiles to mitigate the potential effect of extreme outliers.

Table 2 presents the regression analysis for Equation (6), elucidating the impact of the enactment of cybercrime laws on different indicators of bank stability. The results associated with *Z_SCORE* are outlined in Column (1), while Columns (2) and (3) provide the outcomes for *MES_VAR5%* and *SRISK*, respectively. Across all specifications, the coefficients for *CYBERLAW* are statistically significant and show expected signs, rejecting hypothesis $H_1$, that the enactment of cybercrime laws enhances bank stability.

---

[1] As an alternative analysis, we employ bank- and bank*country- fixed effects and the results, presented in Internet Appendix Table IA2, are qualitatively consistent with the baseline regression.

[Insert Table 2 Here]

In terms of economic significance, the coefficient estimates of 0.025 in Column (1), with *Z_SCORE* as the dependent variable, implies that, on average, individual bank stability increases by 0.69% (0.025/3.579) over the mean value of *Z_SCORE,* following the enactment of cybercrime law. Similarly, when bank stability is measured by *MES_VAR5%* in Column (2), the coefficient estimate is -0.026, indicating that the stability of the banking system experiences a rise of approximately 20.80% (0.026/0.125) after the enactment of the cybercrime law. Similar economic effect is observed for *SRISK*.[2] These results provide robust evidence of the affirmative impact that the enactment of cybercrime laws has on bank stability. In terms of control variables, our results align with previous studies. Higher bank dependence on customer deposits is associated with greater stability and lower bank efficiency destabilizes banks' financial health.

Figure 2 displays the three graphs illustrating the variation in bank stability across three measures - *Z_SCORE, MES_VAR5%,* and *SRISK* over an eleven-year period (i.e., [*t-5* to *t+5*]) centered around the enactment of cybercrime laws.[3] The *Z_SCORE*, representing overall bank stability, shows a steady level before enactment, with a significant increase post-enactment, suggesting improved stability following the enactment of the law. *MES_VAR5%,* which measures the vulnerability of banks to extreme market events, follows a similarly consistent pattern. *MES_VAR5%* shows a decline after t=0, indicating reduced bank vulnerability and risk. Finally, *SRISK*, which measures the expected capital shortfall during crises, shows a noticeable decline post-enactment, implying that banks become less vulnerable to systemic risks following the law's enactment. Collectively, these graphs suggest that cybercrime law enactment has a positive impact on bank stability.

[Insert Figure 2 here]

---

[2] It is important to note that systemic risk is not merely the sum of individual banks' risks. Rather, it depends on how interconnected banks are - that is, the extent to which their exposures are correlated and can lead to simultaneous failures.
[3] We include similar graphs for developed and developing countries in Internet Appendix Figure IA1.

**Addressing endogeneity concerns**

To address potential endogeneity issues, we employ a stacked event-by-event approach to evaluate the impact of cybercrime laws on bank stability. This method addresses concerns raised in recent studies regarding potential biases in treatment effects when using already treated units as comparison units for subsequently treated units (Baker et al., 2022; Cengiz et al., 2019). Additionally, this approach minimizes the risk of confounding omitted variables or unobserved differences between countries influencing the estimate. The enactment of cybercrime laws at the country level is generally independent of individual bank stability policies, making it a suitable basis for a natural experiment. The first difference captures how the change in cybercrime laws impacts bank stability before and after enactment. The second difference assesses how a country's cybercrime laws affect bank stability compared to countries without such laws during a given period. The effect of cybercrime law enactment on bank stability is estimated as the difference between these two differences.

We generate separate datasets for each cybercrime law enactment, excluding countries that have already been treated in subsequent years of enactment. Within these datasets, we employ an 11-year estimation window (from *t-5* to *t+5*) centered around the corresponding enactment year. Subsequently, we stack these event-specific datasets in chronological order to compute the average treatment effect across these events. The outcomes of the stacked event-by-event regressions are presented in Table 3.

[Insert Table 3 Here]

In Table 3, *TREATMENT* is an indicator variable equal to 1 for countries enacting a cybercrime law and zero otherwise during 1990-2021. *POST* is a dummy variable equal to 1 if one to five years post-enactment and zero otherwise. Across all models, we include cohort*country- and cohort *year- fixed effects to account for omitted variable bias.

Consistent with our baseline results, the coefficient of *POST* is positive and statistically significant with *Z_SCORE* as a dependent variable in Column (1). The coefficients of *POST* are negative and statistically significant in Columns (2) and (3) for *MES_VAR5%* and *SRISK*, respectively. This suggests that, on average, banks in countries that have adopted cybercrime laws are more stable. The interaction between *POST*TREATMENT* is statistically significant across all models, indicating that banking system stability becomes more stable after the enactment of cybercrime laws.

Economically, the coefficient of the interaction term *POST\*TREATMENT* is 0.135 in Column (1), indicating that the enactment of cybercrime laws increases bank stability by 13.5%.[4]

The stacked event-by-event analysis requires that treatment and control firms follow parallel trends in the outcome variable before the treatment. To validate the parallel trends assumption, we incorporate lead and lag terms in dynamic difference-in-differences (DiD) regressions (Klasa et al., 2018, Li et al., 2018). Specifically, we generate a new set of indicator variables: $CYBERLAW_{-5}$, $CYBERLAW_{-4}$, $CYBERLAW_{-3}$, $CYBERLAW_{-2}$, $CYBERLAW_{-1}$, $CYBERLAW$, $CYBERLAW_{+1}$, $CYBERLAW_{+2}$, $CYBERLAW_{+3}$, $CYBERLAW_{+4}$, and $CYBERLAW_{+5}$. Variables with positive subscripts reflect whether a country is set to enact the cybercrime law in the next one to five years, while the remaining variables capture whether the law was implemented in the current year or one to five years prior.

We present the results of parallel trend tests in Table 4 and their plots in Figure 3. Table 4 shows that the effect of *CYBERLAW* remains positive and statistically significant. The coefficients for $CYBERLAW_{-5}$ through $CYBERLAW_{-1}$ are close to zero, suggesting no pre-trend bias. In contrast, the coefficient for $CYBERLAW_{+1}$ is positive in Column (1) and of similar magnitude to the main estimates from the baseline results. This indicates that bank stability improves only after the enactment of cybercrime laws, not before, thus confirming the positive impact of cybercrime laws on bank stability.

[Insert Table 4 Here]

[Insert Figure 3 Here]

Next, we employ an instrumental variable (IV) approach to complement the stacked difference-in-differences analysis. We construct an instrument based on the adoption of cybercrime laws by neighboring countries. We define a dummy variable equal to 1 if a country's neighboring countries have adopted cybercrime laws and zero otherwise. This instrument is valid because the adoption of cybercrime laws by neighboring countries can significantly influence a country's own adoption, while being less likely to directly impact the country's banking system stability. By leveraging exogenous variation in cybercrime law adoption from neighboring peer effects, we mitigate concerns that domestic economic shocks or political events drive our findings.

---

[4] Standard staggered difference-in-differences (DiD) estimators can yield biased estimates when treatment timing varies across units, particularly in the presence of heterogeneous treatment effects or dynamic responses (Callaway and Sant'Anna, 2021; Sun and Abraham, 2021). In addition to the stacked regression approach, we further implement the Callaway and Sant'Anna (2021) estimator. This method uses all pre- and post-treatment observations and constructs "clean" control groups comprising firms that are either never treated or not-yet-treated at each treatment date, thereby avoiding contamination from already-treated units. The results, reported in Appendix Table A2, are consistent with those reported in Table 3.

Results from the instrumental variable regression analysis are presented in the Appendix Table A3. The first-stage regression results show that neighboring countries' cybercrime law adoption is positively and significantly associated with a country's own adoption, satisfying the relevance condition. The partial F-statistic indicates that the instrumental variable is not weak. In the second-stage regression, the instrumented *CYBERLAW* variable remains statistically significant and shows the expected signs across all models, supporting the conclusion that cybercrime law adoption has a positive and significant impact on bank stability.

There is a concern that cybercrime law adoption may coincide with broader financial regulatory reforms on capital, liquidity, and governance, meaning observed stability improvements could reflect these changes rather than the cybercrime laws themselves. To address this, we identify major banking regulation reforms that occurred in each country in the same year as cybercrime law enactment using Omori (2022)'s financial reform database. This database covers major financial reforms across 100 countries from 1973 to 2013. For other countries and for the post-2013 sample period (2014–2021), we manually verify major banking reforms using each country's official government website. If a country implemented a significant banking regulation reform in the same year as cybercrime law enactment, we exclude that country from the sample and re-run our baseline regression. Results from Appendix Table A4, Panel A confirm main finding regarding the positive impact of cybercrime law enactment on banking system stability. As an additional robustness test, we remove countries with major banking regulation reforms occurring within the 3-year and 5-year window of the cybercrime law enactment. The results - presented in Appendix Table A4, Panel B and Panel C - are consistent with the baseline results, mitigating concerns that our main findings are driven by financial regulatory changes in periods surrounding the enactment of cybercrime laws.

Finally, we manually examine the historical motivations for cybercrime law enactment across 132 countries. Most adoptions were driven by international commitments, regional harmonization efforts, and responses to major cyber incidents, not by improvements in banking stability or institutional capacity. This evidence mitigates concerns of reverse causality- that cybercrime laws were introduced because stronger financial stability enabled their enactment. Collectively, these results support the causal effect of cybercrime law enactment on bank stability.[5]

**Cross-sectional analysis**

In the cross-sectional analysis, we investigate whether the impact of cybercrime laws on bank stability varies across countries with different regulatory characteristics. First, we assess whether the effect of cybercrime laws on banking system stability is influenced by the severity of these laws. All cybercrime laws in our sample are classified under criminal provisions; therefore, offenders face imprisonment. Stricter penalties should have a stronger deterrent effect, reducing the likelihood of cybercrime. Therefore, we hypothesize that the impact of cybercrime laws on banking system stability will be more significant in countries with stricter laws. We measure severity with *LAW_SEVERITY*, calculated as the natural logarithm of one plus the maximum number of years of imprisonment for a cybercrime offense. We include an interaction term between *LAW_SEVERITY* and *CYBERLAW* in our baseline regression model to examine its effect. The results - presented in Table 5, Panel A - show that the positive relationship between cybercrime laws and bank stability is stronger in countries with more severe penalties, supporting the hypothesis that the effectiveness of cybercrime laws depends on the severity.

Cybercrime legislation also mandates cybersecurity provisions that financial institutions must implement to prevent cyber incidents, with non-compliance incurring significant penalties.[6] We hypothesize that stricter penalties for mishandling customer data increase banks' incentives to establish robust cybersecurity protocols, thereby amplifying the positive impact of cybercrime laws on financial stability. To test this, we collect data on customer data protection regulations for each sample country using the DataGuidance database from OneTrust. We construct the variable *BANK_PENALTY*, representing the natural logarithm of one plus a maximum monetary penalty for non-compliance with data privacy laws, and include the interaction term between *BANK_PENALTY* and *CYBERLAW* in our baseline regression. The results - shown in Table 5, Panel B - indicate that the positive effect of cybercrime laws on bank stability is more pronounced in countries with stricter penalties for data breaches. This finding underscores the importance of holding banks accountable for their cybersecurity practices in strengthening the overall effectiveness of cybercrime laws in promoting financial stability.

---

[5] We also conduct a placebo bootstrap procedure to ensure that our main findings are not driven by random correlations. More specifically, we randomize the enactment year of cybercrime laws for each country and re-run the DiD analysis. This exercise strengthens the causal interpretation of our findings by demonstrating that the observed effects are not caused by spurious correlations or other confounding factors. We perform 1,000 simulations for this analysis. Internet Appendix Figure IA2 shows the distribution of coefficients on the interaction term *POST\*TREATMENT* obtained from the 1,000 simulations for each bank stability measure. The maximum value of the simulated *POST\*TREATMENT* for the DiD analysis involving *Z_SCORE* is 0.036, which is nearly four times less than the genuine *POST\*TREATMENT* coefficient (0.135) shown in Table 3, Column (1). Moreover, the standard deviation of the bootstrap distribution for simulated *POST\*TREATMENT* coefficients is 0.010, which is considerably smaller than the standard error of *POST\*TREATMENT* (0.051) in Table 3 Column (1). This statistic implies that the standard error on the *POST\*TREATMENT* coefficient estimate is conservative. We obtain similar evidence for other bank stability measures. Taken together, results from the placebo bootstrap analysis suggest that the significant estimates on *POST\*TREATMENT* in the DiD regressions are derived from the genuine relationship between cybercrime law enactment and bank stability rather than spurious correlations.

[6] For example, in the United Kingdom, the legal framework for both cybercrime and cybersecurity are closely interconnected through key pieces of legislation, including the Computer Misuse Act (1990). This legislation plays a dual role in both protecting individuals and enhancing cybersecurity. Offences under the Computer Misuse Act (1990) can arise from unauthorized access to personal data, which constitutes both a cybercrime and a breach of cybersecurity obligations due to inadequate measures.

Recent cross-border cyberattacks underscore the need for international legal cooperation; Iran's DDoS attacks on US financial institutions and North Korea's cyber heists show how threats traverse borders. Therefore, we investigate if international cooperation through extradition treaties can moderate the effect of cybercrime laws. A bilateral extradition treaty enables one country to surrender a person to another for prosecution (e.g., under the US-UK treaty, the UK may request the surrender of a US resident who committed a cybercrime against UK firms). Such cooperation is likely to inhibit a person's intention to commit a crime in another country, therefore, we expect that the effectiveness of a cybercrime law enacted by a country depends on the extent to which the country has bilateral extradition treaties with other countries. To test this, we manually collect data on extradition treaties for each country in our sample. We construct the variable *EXTRADITION*, which is the natural logarithm of one plus the number of extradition treaties a country has each year. We include the interaction term between *EXTRADITION* and *CYBERLAW* in the baseline regression model. As shown in Table 5 Panel C, the positive effect of cybercrime law enactment on banking system stability is stronger among countries with more extradition treaties, supporting the notion that international cooperation through extradition treaties enhances the effectiveness of cybercrime laws.

Lastly, the rule of law in a country affects how laws are applied and enforced. In a country where the rule of law is strong, cybercrime laws are more likely to be applied consistently and fairly, enhancing their effectiveness and fostering more secure banking sectors, as clients trust that violations will be penalized and threats mitigated. To test this, we include the interaction term between *RULE OF LAW* and *CYBERLAW* in the baseline regression and present their results in Table 5, Panel D. We obtain a positive and statistically significant coefficient for the interaction term *RULE OF LAW * CYBERLAW* in Column (1) and negative and statistically significant coefficients in Columns (2) and (3). This suggests that the impact of cybercrime regulations is more pronounced in countries whose legal system is robust in enforcing their laws.

[Insert Table 5 here]

**ECONOMIC CHANNELS**

**Funding liquidity channel**

Cybercrime laws may significantly affect bank stability by influencing funding liquidity. Stringent regulations help institutions protect against cyber-attacks, safeguard customer data, prevent fraud, and maintain transaction integrity. This heightened security can enhance customer confidence, leading to increased bank deposits and wholesale funding while reducing interbank funding risks (Adelmann et al., 2020), thereby strengthening the bank's funding liquidity position. However, the issue of moral hazard may arise if banks become overly dependent on the protection provided by these laws. In a regulatory environment where cyber risks are perceived as mitigated by legislation, banks may feel less pressure to maintain robust internal cybersecurity measures. This potential moral hazard could undermine bank funding liquidity, as stakeholders may become reluctant to provide funding (e.g., through deposits or lending) to institutions they perceive as overly reliant on regulatory safeguards.

To empirically explore this funding liquidity channel, we employ a system of equations using path analysis, presented as follows:

$$Bank\ Stability_{i,c,t} = \alpha + \beta_1 CYBERLAW_{c,t} + \beta_2 Funding\ Liquidity_{i,c,t} + \beta_3 Bank\ Controls_{i,c,t} + \beta_4 Macro\ Controls_{c,t} + v_c + \mu_t + \varepsilon_{i,c,t} \quad (7)$$

$$Funding\ Liquidity_{i,c,t} = \alpha + \gamma_1 CYBERLAW_{c,t} + \gamma_2 Bank\ Controls_{i,c,t} + \gamma_3 Macro\ Controls_{c,t} + v_c + \mu_t + \varepsilon_{i,c,t} \quad (8)$$

Funding liquidity refers to banks' ability to access stable funding sources, including deposits, wholesale funding, and interbank funding (Altavilla et al., 2025). To capture changes in deposits and overall funding around cybercrime law enactment, we construct four measures: 1) a change in total bank funding (∆*TOTALFUNDING*), 2) a change in the net stable funding ratio (∆*NSFR*), 3) a change in deposits (*ΔDEPOSIT*), and 4) a change in the deposit-to-assets ratio (*ΔDEPTOASSET*). ∆*TOTALFUNDING* is the difference between the current period's total funding and the previous period's total funding, divided by the previous period's funding, and then multiplied by 100. ∆*NSFR* is the ratio of the percentage change in available stable funding to the percentage change in its required stable funding between two consecutive periods. *ΔDEPOSIT* is the difference between the current period's deposit levels and the previous period's deposit levels, expressed as a percentage change. *ΔDEPTOASSET* is the difference between the deposit-to-assets ratio in the current period and the previous period, expressed as a percentage change. These variables allow us to assess whether cybercrime law enactment influences availability as well as changes in bank funding sources.

In Equation (7), the direct path from *CYBERLAW* to *Bank Stability* is captured as $β_1$. The magnitude of the path from *Funding Liquidity* to *Bank Stability* is denoted as $β_2$. In Equation (8), the path coefficient $γ_1$ indicates the magnitude of the path from *CYBERLAW* to *Funding Liquidity*. The interaction $γ_1 \times β_2$ represents the total magnitude of the indirect effect of *CYBERLAW* on *Bank Stability* channeling through *Funding Liquidity*. We illustrate this association in Figure 4, below.

[Insert Figure 4 here]

We present the estimation results of Equations (7) and (8) in Table 6. Panels A, B, C, and D report results for ∆*TOTALFUNDING*, ∆*NSFR, ΔDEPOSIT,* and *ΔDEPTOASSET,* respectively. Providing support to baseline findings, the path analysis results indicate a positive association between the enactment of cybercrime laws and bank stability. In addition, cybercrime law enactment is positively related to bank funding liquidity. The path analysis shows that the funding liquidity channel can explain up to 15.72% of the total effect of cybercrime law enactment on bank stability. Collectively, these findings lend credence to the notion that cybercrime laws can bolster bank stability by enhancing funding liquidity.[7]

[Insert Table 6 here]

---

[7] As a robustness test, we utilize the volatility of the deposit-to-assets ratio to capture the stability of deposit funding and document similar results in Internet Appendix Table IA3.

**Operational risk channel**

The enactment of cybercrime laws can impact bank stability through operational risk. Prior studies (e.g., Berger et al., 2022; Curti et al., 2022) show that operational risk can undermine bank stability. In discouraging one's intent to commit illegal cyber activities, cybercrime laws can reduce the risk of cyberattacks. Cybercrime laws also encourage banks to identify vulnerabilities, adapt to evolving cyber threats, frequently update their security systems, and maintain the integrity of their digital infrastructure, minimizing operational risk. In the event of a cyberattack, the established protocols and updated security systems facilitate a swift and effective response, mitigating the potential damage and associated financial losses. Therefore, cybercrime laws can enhance the overall stability of banks through the reduction of operational risk. However, perceived protection may induce complacency, leading banks to pay less attention to cybersecurity and increasing operational risk.

Accordingly, we test the operational risk channel by conducting a path analysis, with the corresponding path model mathematically expressed as follows:

$$Bank\ Stability_{i,c,t} = \alpha + \beta_1 CYBERLAW_{c,t} + \beta_2 Bank\ Operational\ Risk_{i,c,t} +$$
$$\beta_3 Bank\ Controls_{i,c,t} + \beta_4 Macro\ Controls_{c,t} + v_c + \mu_t + \varepsilon_{i,c,t} \qquad (9)$$

$$Bank\ Operational\ Risk_{i,c,t} = \alpha + \gamma_1 CYBERLAW_{c,t} + \gamma_2 Bank\ Controls_{i,c,t} +$$
$$\gamma_3 Macro\ Controls_{c,t} + v_c + \mu_t + \varepsilon_{i,c,t} \qquad (10)$$

Operational risk in banking is defined as the risk stemming from failed systems, people, internal processes, and external factors (BCBS, 2004). We estimate three alternative proxies to capture different dimensions of banks' exposure to operational risk: (1) non-interest expense to total bank operational expenses, (2) operational risk density computed non-interest expense to total assets, and (3) operational risk efficiency measured as the ratio of non-interest expense to total revenues.

In Equations (9) and (10), *β₁* captures the direct effect of *CYBERLAW* on *Bank Stability*, and the coefficient *γ₁* represents the impact of *CYBERLAW* on *Bank Operational Risk*. The magnitude of the effect of *Bank Operational Risk* on *Bank Stability* is captured by *β₂*, and the total magnitude of the indirect effect of *CYBERLAW* on *Bank Stability* through *Bank Operational Risk* is captured by *γ₁ x β₂*. We demonstrate this relationship in Figure 5.

[Insert Figure 5 here]

The estimation results for Equations (9) and (10) are presented in Table 7. Panel A reports overall operational risk, Panel B operational risk density, and Panel C operational risk efficiency. Our findings provide further confirmation of the baseline analysis, revealing a positive association between the enactment of cybercrime laws and bank stability. Results across different measures of bank operational risk consistently indicate that bank operational risk is negatively associated with the enactment of cybercrime laws. In addition, there is significant evidence suggesting bank operational risk serves as a mediator in the relationship between cybercrime law enactment and bank stability. For example, as shown in Table 7, Panel B, the bank's operational risk density channel accounts for 23.05%, 8.80%, and 6.33% of the total effect of cybercrime law enactment on bank stability, measured by *Z_SCORE*, *MES_VAR5%*, and *SRISK*, respectively.

Our results suggest that the operational risk channel predominantly mediates the impact of cybercrime laws on individual bank risk (*Z_SCORE*) compared to its effect on systemic risk within the banking sector (*MES_VAR5%* and *SRISK*). This finding implies that the primary benefits of cybercrime laws in mitigating operational risk are observed at the individual bank level, with a relatively smaller yet still significant effect on broader financial stability.

[Insert Table 7 here]

**Effectiveness of Cybercrime Laws**

As our main hypotheses are built upon the notion that cybercrime laws are effective in mitigating the risk of cyber incidents, we further conduct an empirical test to provide direct evidence regarding the effect of cybercrime laws on cyber outcomes. To test this, we gather data on cyber incidents around the world from the University of Maryland–Center for International and Security Studies at Maryland (CISSM) cyber-attacks database. We construct the variable *CYBER_INCIDENTS* as the natural logarithm of one plus the number of cyber incidents in a particular year for each country. We also proxy for the severity of cyber-attacks through the monetary damage caused by such attacks. Accordingly, we compute the variable *INCIDENT_LOSS* as the natural logarithm of one plus the total dollar loss due to cyber-attacks faced by a country in a given year. We then regress each of *CYBER_INCIDENTS* and *INCIDENT_LOSS* on *CYBERLAW* along with a set of country-level control variables. We report the results of these regressions in Table 8. We document evidence showing a decrease in the number of cyber incidents and their severity after the enactment of cybercrime laws. These findings highlight the direct role of cybercrime laws in reducing the risk of cyber offenses.

[Insert Table 8 here]

## ADDITIONAL ANALYSES AND ROBUSTNESS TESTS

**Country legal origin**

We further examine whether the impact of cybercrime laws on bank stability varies by a country's legal origin. Following Liang and Renneboog (2017), we classify legal origin into five categories: English common law, French civil law, German civil law, Scandinavian civil law, and Socialist law. Results from Internet Appendix Table IA4 suggest that cybercrime laws negatively impact bank stability in German civil law countries. This is possibly due to a more rigid legal framework that, while aiming to enhance security, could limit banks' ability to quickly adapt to evolving cyber threats. Unlike in common law countries, where laws are more flexible and dispute-focused, civil law nations may struggle to align their cybercrime regulations with the dynamic nature of cyber risks. Results for countries with a Socialist origin remain inconclusive.

**Fintech development**

We propose that the effect of cybercrime laws on bank stability also varies with a country's level of financial development. Technological adoption in financial systems differs across countries (Kumar et al., 2021; Luo, 2021), affecting how banks manage cyber threats. In more advanced financial systems, where IT infrastructure is integral to complex financial transactions, institutions require greater security measures. Thus, we expect cybercrime laws to be more effective in banks operating in countries with more fintech development. Obtaining financial technology development data from the World Bank, we use the metric "made or received a digital payment," which reflects the level of digital financial transaction adoption. As shown in Internet Appendix Table IA5, cybercrime laws have a more pronounced effect on bank stability in countries with higher financial technology development.

**Moral hazard**

We also explore whether cybercrime laws may lead to moral hazard, where banks become overly reliant on the protection provided by these laws. We proxy for the potential risk of moral hazard using deposit insurance. Many countries implement explicit deposit insurance as a mechanism to protect depositors in the event of bank failures (Cull et al., 2005). While deposit insurance is intended to build trust in the banking system and thereby promotes financial stability, its negative byproduct is the emergence of moral hazard, as banks have incentives to engage in riskier activities than they otherwise would (Grossman, 1992; Wheelock and Kumbhakar, 1995). For each country, we obtain data on explicit deposit insurance from the World Bank's Deposit Insurance Database compiled by Demirgüç-Kunt et al. (2013). The results in Table IA6 reveal that moral hazard is unlikely to drive the relationship between cybercrime laws and bank stability.

One possible explanation is that the explicit deposit insurance may be an imperfect proxy for bank moral hazard. Banks' risk-taking incentives are fundamentally driven by structural features such as high leverage and limited liability, which persist irrespective of deposit insurance coverage (John et al., 2000). While deposit insurance may amplify these incentives by insulating depositors, it is not their root cause. Moreover, the design of deposit insurance schemes plays a critical role in determining their effect on bank risk-taking (Gropp and Vesala, 2004). Well-structured programs with clearly defined coverage limit and risk-based premiums can mitigate moral hazard by credibly limiting the government safety net and encouraging monitoring by uninsured creditors. Simply controlling for whether a country has explicit deposit insurance might not fully capture this variation in design and its potential disciplining effect. Overall, our findings do not support the view that cybercrime laws promote excessive risk-taking or bank complacency. Rather, the evidence is more consistent with the interpretation that such laws reduce operational risk and improve bank funding liquidity, thereby enhancing banking system stability. However, we do not rule out the existence of moral hazard that may be less easily captured by standard proxies like explicit deposit insurance. Investigating these subtler dynamics represents a valuable direction for future research.

**Different types of cyber regulations**

There are various types of cyber regulations, including e-transactions law, data protection and privacy law, and online consumer protection law. These legislations each serve distinct purposes and may influence the association between the enactment of cybercrime law and bank stability. As shown in Table IA7, the presence of other cyber laws, such as data privacy, e-transactions, and consumer protection regulations, can moderate the impact of cybercrime law enactment on banking system stability.

**Banking market structure and regulations**

The relationship between cybercrime law and bank stability may vary with banking industry concentration. Higher concentration, driven by digital economies of scale or other market forces, creates single or near-single points of failure, increasing susceptibility to cyber shocks. Digital consolidation in core services (treasury clearing, settlement, and cloud services) further heightens risk.[8] Thus, the enactment of cybercrime laws becomes crucial in concentrated banking sectors which are more exposed to cyber threats, since successful cyberattacks could have far-reaching consequences for the stability and integrity of the entire financial system. The results in Table IA8 suggest that the positive impact of cybercrime law is more pronounced for banks in countries with increased market concentration.

We further investigate the role of bank ownership. Results in Table IA9 suggest that while cybercrime laws improve overall stability, this effect is less pronounced for state-owned banks. A possible explanation is that state-owned banks are perceived as more stable due to implicit government guarantees, which reduce their vulnerability to short-term financial pressures (Borisova and Megginson, 2011; Boubakri et al., 2018). Additionally, state-owned banks prioritize public welfare over profit maximization, leading to conservative risk management that further shields them from potential shocks.

We also examine the role of foreign ownership. Results in Table IA10 show that cybercrime laws have a stronger positive impact on banks with higher foreign ownership. Foreign-owned banks often leverage advanced technologies and have access to a skilled workforce, allowing them to implement more sophisticated cybersecurity measures (Berger et al., 2000).

---

[8] For instance, while central clearing counterparties (CCPs), which represent increased concentration, have contributed to reducing vulnerabilities in the aftermath of the 2008-09 financial crisis by enhancing transparency and netting, they also introduce new cyber risks. This is because CCPs concentrate financial activities within a single entity, potentially magnifying the impact of cyberattacks if adequate investment in cyber resilience is lacking.

Lastly, we analyze the effect of various banking regulations, including those governing bank activities, capital requirements, and supervisory power (Barth et al., 2004). Table IA11 shows that in countries with stricter banking regulations, the effect of cybercrime laws on bank stability is less pronounced.

**Bank size**

We also assess whether the effect of cybercrime laws on bank stability varies by bank size. Table IA12 suggests that the positive effect of cybercrime laws is more significant for larger banks. This finding is consistent with the notion that larger banks have greater resources to invest in cybersecurity, recognizing the significant reputational and monetary risks of cyberattacks (Crisanto and Prenio, 2017).

**Banking sector development**

We further examine whether banking sector development impacts the association between cybercrime laws and banking stability. As shown in Table IA13, the positive effect of cybercrime laws is stronger in countries with more developed banking sectors. This aligns with Allen et al. (2021), who show that financial development enhances systemic resilience by improving efficiency and risk management, which can amplify the stabilizing impact of cybercrime laws.

**Change analysis**

To control for potential omitted variables, we conduct change regressions by re-estimating all models in Equation (6) using annual changes in firm-level variables (Bates et al., 2009). The dependent variable is the change in bank stability from year $t$ to $t+1$, while independent variables reflect changes in bank-level and country-level controls from year $t-1$ to $t$. Table IA14 confirms a positive association between cybercrime law enactment and changes in bank stability, lending support to the main findings.

**Additional robustness tests**

To further ensure the robustness of our results, we perform several additional tests. First, we exclude U.S. banks, given their large representation in our sample. Second, we remove observations from the Global Financial Crisis (2007-2009) to control for the unique effects of this period on bank stability. Third, we control for various regulatory determinants of banking performance, including bank activity restrictions, bank capital regulation, and supervisory power (Barth et al., 2004; Cope et al., 2012). Fourth, we exclude countries that enacted significant banking regulation reforms within a five-year window of the enactment of the cybercrime laws and re-perform the DiD analysis. Finally, we exclude countries that enacted cybercrime laws in direct response to major cyber incidents and re-estimate our baseline regressions. Results from these robustness tests, presented in Table IA15, Table IA16, Table IA17, Table IA18, and Table IA19 confirm our main findings, highlighting the important role of cybercrime laws in enhancing bank stability.[9]

## CONCLUSION

As financial intermediation is crucial to economic growth, understanding what affects banking system stability is of particular interest to academics, practitioners, and regulators. In this paper, we analyze the impact of cybercrime law enactments on banking system stability using data for 132 countries. We document the significant favorable effect of cybercrime regulations on bank stability, highlighting their importance in strengthening financial stability. This result is robust across extensive tests.

We further find that this positive association is stronger in countries with heavier punishments for unlawful cyber activities; in regimes that hold banks accountable for their cybersecurity practices, with greater international legal cooperation via extradition treaties; and where the rule of law ensures consistent enforcement.

We further examine two channels: funding liquidity and operational risk. We show that cybercrime laws bolster depositors' and investors' confidence, increasing deposits and stabilizing funding sources. In addition, by prosecuting cyber intruders and demanding banks implement robust cybersecurity programs, cybercrime laws play an important role in mitigating the cyber risk facing the banking sector. Collectively, funding liquidity enhancement and operational risk reduction are the two main mechanisms through which cybercrime laws can bolster the banking system's stability.

---

[9] In untabulated analysis, we utilize the marginal expected shortfall calculated at the 1$^{st}$ percentile of value at risk (*MES_VAR1%*) as an alternative measure to *MES_VAR5%* and document consistent evidence regarding the impact of cybercrime law enactment on bank stability.

This study provides strong policy implications. Our paper suggests that international regulatory bodies (e.g., the Financial Stability Board, the Committee on Payments and Market Infrastructure, and the Basel Committee) should encourage their member states to adopt cybercrime regulations to mitigate the exposure of the banking system to the risk of cyberattacks.[10] While Basel III emphasizes capital adequacy, liquidity requirements, and risk management to enhance financial system resilience, these measures may be insufficient in addressing the unique risks posed by cyber incidents. For instance, while capital and liquidity can provide financial resources to respond to a cyber incident, they do not necessarily accelerate the recovery of compromised systems or data. Given the growing complexity of cyber risks, regulatory frameworks must evolve beyond traditional financial safeguards to incorporate robust cybersecurity measures.

Given that cyber threats transcend borders, cybercrime laws should be treated as an integral component of international business regulation. Isolated efforts are unlikely to address systemic cyber risk. International collaboration among law enforcement is pivotal, and, due to global financial interdependence, a cyber incident in one country can spill over to others. Our findings imply that a country's adoption of cybercrime regulation can contribute to global banking stability. We call for future research on cross-border effects of cybercrime laws and on how such initiatives can complement Basel III to address systemic cyber risks.

---

[10] While traditional regulatory tools—such as capital and liquidity requirements—may be effective in some instances, cyberattacks often necessitate additional measures, such as instituting bank holidays for system restoration or temporarily waiving liquidity requirements (Eisenbach et al., 2022). The asymmetrical information stemming from cyber incidents also emphasizes the importance of proactive policy interventions. For example, disclosing even minor cyber events to regulators or sharing threat assessments and contingency plans among banks can bolster resilience by minimizing uncertainty and fostering collective learning. Addressing coordination breakdowns following cyber incidents is equally vital. Approaches such as identifying affected banks or certifying key banks as unaffected can help diminish uncertainty. Nevertheless, reliance solely on information disclosure policies may fall short in alleviating coordination failures when policies are internally determined (Angeletos et al., 2006). Thus, a multifaceted approach integrating information disclosure with proactive policy measures is imperative to effectively mitigate the systemic risks posed by cyber events. The role of central banks and other entities in providing dedicated backup facilities in critical markets is another potential avenue (Duffie and Younger, 2019).

**REFERENCES**

Acharya, V. V. (2009). A theory of systemic risk and design of prudential bank regulation. *Journal of Financial Stability*, *5*(3), 224–255.

Acharya, V. V., Pedersen, L. H., Philippon, T., & Richardson, M. (2017). Measuring systemic risk. *Review of Financial Studies, 30*(1), 2-47.

Acharya, V., Engle, R., & Richardson, M. (2012). Capital shortfall: A new approach to ranking and regulating systemic risks. *American Economic Review, 102*(3), 59-64.

Adelmann, F., Elliott, M. J. A., Ergen, I., Gaidosch, T., Jenkinson, N., Khiaonarong, M. T., Wilson, C. (2020). Cyber risk and financial stability: It's a small world after all: *International Monetary Fund*.

Adrian, T., & Brunnermeier, M. K. (2016). CoVaR. *American Economic Review*, *106*(7), 1705–1741.

Allen, F., Carletti, E., Cull, R., Qian, J. Q. J., Senbet, L., & Valenzuela, P. (2021). Improving access to banking: Evidence from Kenya. *Review of Finance, 25*(2), 403-447.

Allen, F., & Gale, D. (2000). Financial contagion. *The Journal of Political Economy, 108*(1), 1–33.

Altavilla, C., Boucinha, M., Burlon, L., Giannetti, M., & Schumacher, J. (2025). Central bank liquidity reallocation and bank lending: Evidence from the tiering system. *Journal of Financial Economics*, *168*.

Altermatt, L., van Buggenum, H., & Voellmy, L. (2024). Systemic bank runs without aggregate risk: How a misallocation of liquidity may trigger a solvency crisis. *Journal of Financial Economics*, *161*.

Angeletos, G., Hellwig, C., & Pavan, A. (2006). Signaling in a global game: Coordination and policy traps. *Journal of Political Economy, 114*(3), 452–484.

Baker, A. C., Larcker, D. F., & Wang, C. C. (2022). How much should we trust staggered difference-in-differences estimates? *Journal of Financial Economics, 144*(2), 370-395.

Barth, J. R., Caprio, G., & Levine, R. (2004). Bank regulation and supervision: what works best? *Journal of Financial Intermediation, 13*(2), 205–248.

Barton, C., Anderson, R., Levi, M., Böhme, R., van Eeten, M., Moore, T., Savage, S., & Clayton, R. (2013). Measuring the cost of cybercrime. In *The Economics of Information Security and Privacy, Springer*, 265–300.

Bates, T. W., Kahle, K. M., & Stulz, R. M. (2009). Why do U.S. firms hold so much more cash than they used to? *Journal of Finance*, *64*(5), 1985–2021.

BCBS. (2004). Basel II: International convergence of capital measurement and capital standards: A revised framework. *Basel Committee Publications*.

Berger, A. N., Curti, F., Lazaryan, N., Mihov, A., & Roman, R.A. (2023). Climate risks in the U.S. banking sector ; Evidence from operational losses and extreme storms. *Federal Reserve Bank Philadelphia Working Paper*.

Berger, A. N., Curti, F., Mihov, A., & Sedunov, J. (2022). Operational risk is more systemic than you think: Evidence from US bank holding companies. *Journal of Banking & Finance*, *143*, 106619.

Berger, A. N., & DeYoung, R. (1997). Problem loans and cost efficiency in commercial banks. *Journal of Banking & Finance, 21*(6), 849-870.

Berger, A. N., DeYoung, R., Genay, H., & Udell, G. F. (2000). Globalization of financial institutions: Evidence from cross-border banking performance. *Brookings-Wharton Papers on Financial Services, 2000*(1), 23–120.

Berger, A. N., Guedhami, O., Kirimhan, D., Li, X., & Zhao, D. (2024). Universal banking powers and liquidity creation. *Journal of International Business Studies*, *55(6)*, 764–781.

Borisova, G., & Megginson, W. L. (2011). Does government ownership affect the cost of debt? Evidence from Privatization. *Review of Financial Studies, 24*(8), 2693–2737.

Boubakri, N., El Ghoul, S., Guedhami, O., & Megginson, W. L. (2018). The market value of government ownership. *Journal of Corporate Finance, 50*, 44–65.

Brownlees, C., & Engle, R. F. (2017). SRISK: A conditional capital shortfall measure of systemic risk. *Review of Financial Studies, 30*(1), 48-79.

Callaway, B., & Sant'Anna, P. H. (2021). Difference-in-differences with multiple time periods. *Journal of Econometrics*, *225*(2), 200-230.

Caballero, R. J., & Simsek, A. (2013). Fire sales in a model of complexity. *Journal of Finance, 68*(6), 2549–2587.

Cengiz, D., Dube, A., Lindner, A., & Zipperer, B. (2019). The effect of minimum wages on low-wage jobs. *Quarterly Journal of Economics, 134*(3), 1405-1454.

Chen, T.-Y., Chen, Y.-C., & Hung, M. (2022). Uneven regulatory playing field and bank transparency abroad. *Journal of International Business Studies*, *53(3)*, 379–404.

Chen, W., Li, X., Wu, H., & Zhang, L. (2024). The impact of managerial myopia on cybersecurity: Evidence from data breaches. *Journal of Banking & Finance*, *166*.

Cope, E. W., Piche, M. T., & Walter, J. S. (2012). Macroenvironmental determinants of operational loss severity. *Journal of Banking & Finance, 36*(5), 1362-1380.

Crisanto, J., & Prenio, J. (2017). Regulatory approaches to enhance banks' cyber-security frameworks. *Bank for International Settlements Publications*.

Crosignani, M., Macchiavelli, M., & Silva, A. F. (2023). Pirates without borders: The propagation of cyberattacks through firms' supply chains. *Journal of Financial Economics, 147*(2), 432–448.

Cull, R., Senbet, L. W., & Sorge, M. (2005). Deposit insurance and financial development. *Journal of Money, Credit and Banking, 37*(1), 43-82.

Curti, F., Frame, W. S., & Mihov, A. (2022). Are the largest banking organizations operationally more risky?. *Journal of Money, Credit and Banking*, *54*(5), 1223-1259.

Dam, L., & Koetter, M. (2012). Bank bailouts and moral hazard: Evidence from Germany. *The Review of Financial Studies*, *25*(8), 2343-2380.

Demirgüç-Kunt, A., Kane, E., & Laeven, L. (2013). Deposit Insurance Database. *IMF Working Paper*.

Dietrich, A., & Wanzenried, G. (2011). Determinants of bank profitability before and during the crisis: Evidence from Switzerland. *Journal of International Financial Markets, Institutions & Money, 21*(3), 307-327.

Duffie, D., Younger, J. (2019). Cyber runs. *Hutchins Center on Fiscal and Monetary Policy Working Paper No. 51*. Washington, DC: Brookings Institution.

Eisenbach, T. M., Kovner, A., & Lee, M. J. (2022). Cyber risk and the U.S. financial system: A pre-mortem analysis. *Journal of Financial Economics, 145*(3), 802-826.

Erol, S., & Vohra, R. (2022). Network formation and systemic risk. *European Economic Review, 148*, 104213.

Florackis, C., Louca, C., Michaely, R., Weber, M. Cybersecurity risk (2022). *Review of Financial Studies*. *36*(1), 351-407.

Goetz, M. R. (2018). Competition and bank stability. *Journal of Financial Intermediation*, *35*, 57–69.

Gropp, R., & Vesala, J. (2004). Deposit Insurance, Moral Hazard and Market Monitoring. *European Finance Review*, *8*(4), 571–602.

Grossman, R. S. (1992). Deposit insurance, regulation, and moral hazard in the thrift industry: Evidence from the 1930's. *American Economic Review, 82*(4), 800–821.

Jamilov, R., Rey, H., & Tahoun, A. (2023). The Anatomy of Cyber Risk. *NBER Working Paper Series*.

Jia, J., Jin, G. Z., & Wagman, L. (2018). The short-run effects of GDPR on technology venture investment. *National Bureau of Economic Research Working Paper* No. 25248.

John, G., Phil, M., & John, O. S. W. (2004). Dynamics of growth and profitability in banking. *Journal of Money, Credit and Banking, 36*(6), 1069-1090.

John, K., Saunders, A., & Senbet, L. W. (2000). A theory of bank regulation and management compensation. *Review of Financial Studies, 13*(1), 95–125.

Kamiya, S., Kang, J.-K., Kim, J., Milidonis, A., & Stulz, R. M. (2021). Risk management, firm reputation, and the impact of successful cyberattacks on target firms. *Journal of Financial Economics, 139*(3), 719-749.

Kashyap, A. K., & Wetherilt, A. (2019). Some principles for regulating cyber risk. *AEA Papers and Proceedings, 109*, 482-487.

Kaufman, D., & Kraay, A. (2019). World governance indicators (WGI). *Brookings Institution, World Bank Development Economics Research Group*.

Khan, M. S., Scheule, H., & Wu, E. (2017). Funding liquidity and bank risk taking. *Journal of Banking & Finance, 82*, 203-216.

Klasa, S., Ortiz-Molina, H., Serfling, M., & Srinivasan, S. (2018). Protection of trade secrets and capital structure decisions. *Journal of Financial Economics, 128*(2), 266–286.

Kumar, V., Nim, N., & Agarwal, A. (2021). Platform-based mobile payments adoption in emerging and developed countries: Role of country-level heterogeneity and network effects. *Journal of International Business Studies, 52*(8), 1529–1558.

Levi, M. (2017). Assessing the trends, scale and nature of economic cybercrimes: overview and Issues: In *Crime, Law, and Social Change*, *67*(1), 3–20.

Li, Y., Lin, Y., & Zhang, L. (2018). Trade secrets law and corporate disclosure: Causal evidence on the proprietary cost hypothesis. *Journal of Accounting Research, 56* (2018), 265–308.

Liang, H., & Renneboog, L. (2017). On the foundations of corporate social responsibility. *The Journal of Finance, 72*(2), 853–910.

Luo, Y. (2021). A general framework of digitization risks in international business. *Journal of International Business Studies, 53*, 344–361.

Madan, S., Savani, K., & Katsikeas, C. S. (2023). Privacy please: Power distance and people's responses to data breaches across countries. *Journal of International Business Studies, 54*(4), 731–754.

Micco, A., Panizza, U., & Yañez, M. (2007). Bank ownership and performance. Does politics matter? *Journal of Banking & Finance, 31*(1), 219-241.

Nazil, S.A, Karagozoglu,A. & Zhou, T. (2021). Firm-level cybersecurity risk and idiosyncratic volatility, *Journal of Portfolio Management, 47*(6), 110-140.

North, D. C. (1990). *Institutions, institutional change and economic performance*. Cambridge: Cambridge University Press.

Omori, S. (2022). Introducing the revised and updated financial reform database. *Journal of Financial Regulation*, *8*(2), 230-240.

Raz, A. F., McGowan, D., & Zhao, T. (2022). The dark side of liquidity regulation: Bank opacity and funding liquidity risk. *Journal of Financial Intermediation*, *52*.

Scott, W. R. (1995). *Institutions and organizations. Foundations for organizational science*. London: Sage.

Shim, J. (2019). Loan portfolio diversification, market structure and bank stability. *Journal of Banking & Finance, 104*, 103-115.

Sun, L., & Abraham, S. (2021). Estimating dynamic treatment effects in event studies with heterogeneous treatment effects. *Journal of Econometrics*, *225*(2), 175-199.

Thakor, A. V. (2023). Finance research: What are the new frontiers? *Financial Review,* 1-10.

United Nations Office on Drugs and Crime. (2013). Comprehensive Study on Cybercrime.

Wheelock, D. C., & Kumbhakar, S. C. (1995). Which banks choose deposit insurance? Evidence of adverse selection and moral hazard in a voluntary insurance system. *Journal of Money, Credit and Banking, 27*(1), 186–201.

**Figure 1: Cybercrime Legislation Worldwide**

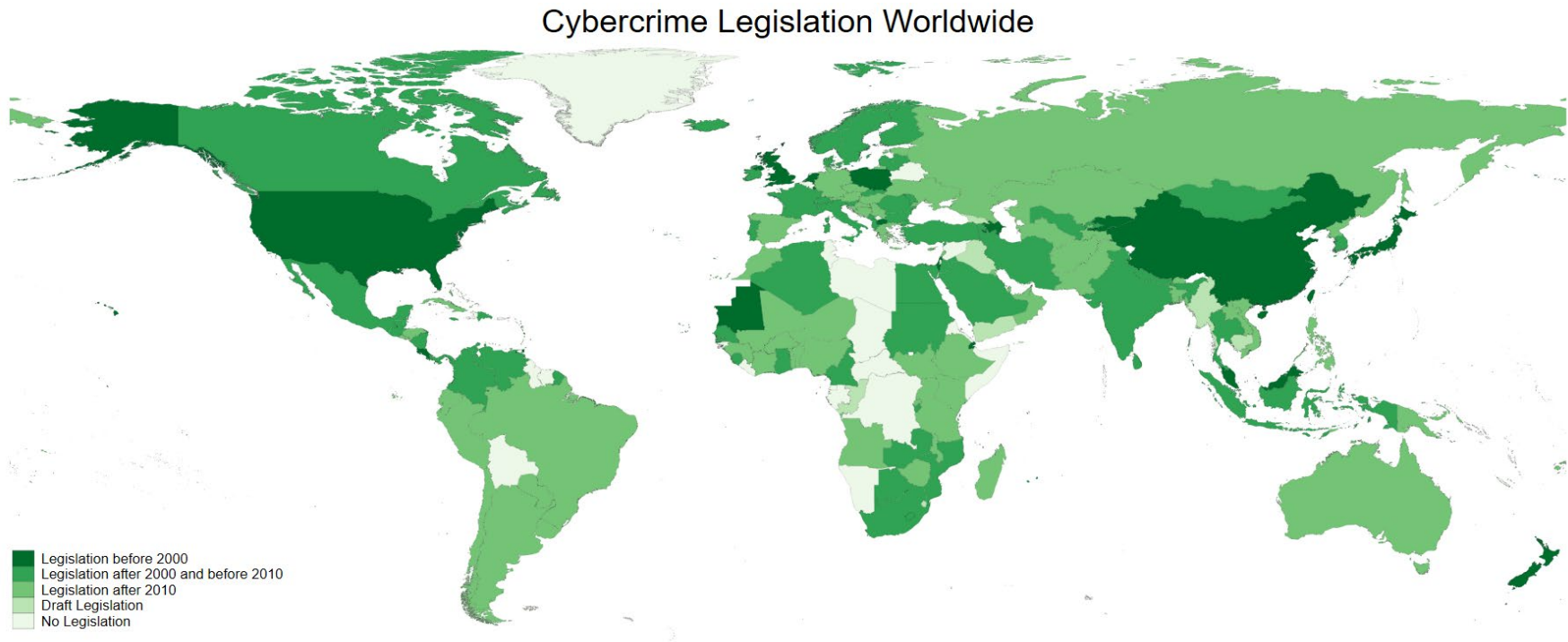

This figure shows the status of cybercrime law enactments around the world as of December 2021.

**Figure 2: Cybercrime Laws and Bank Stability**

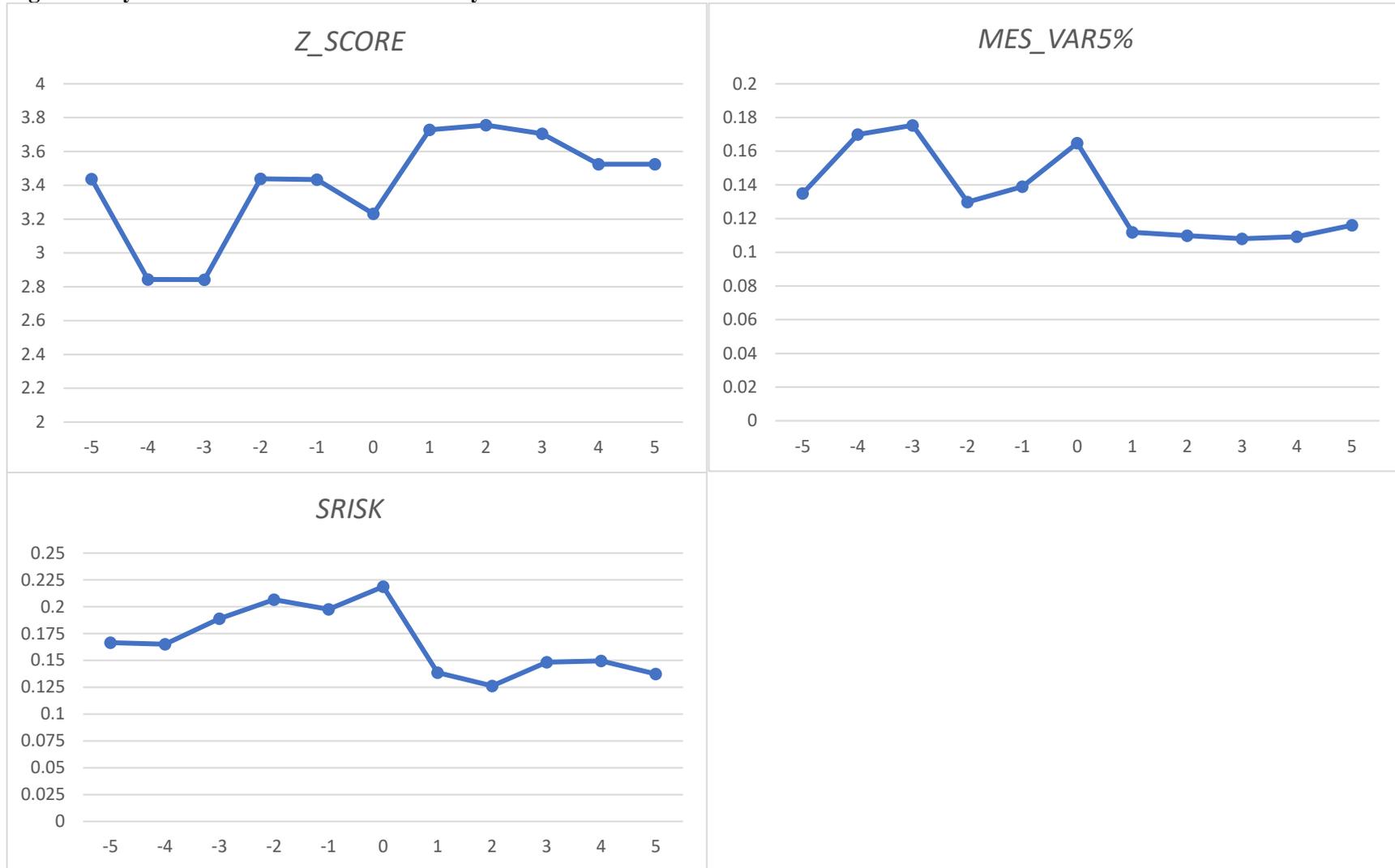

Note: This figure illustrates the variations across different measures of bank stability over a 11-year period surrounding the enactment of cybercrime laws. The figures are drawn from our baseline sample.

**Figure 3: Parallel Trend Tests**

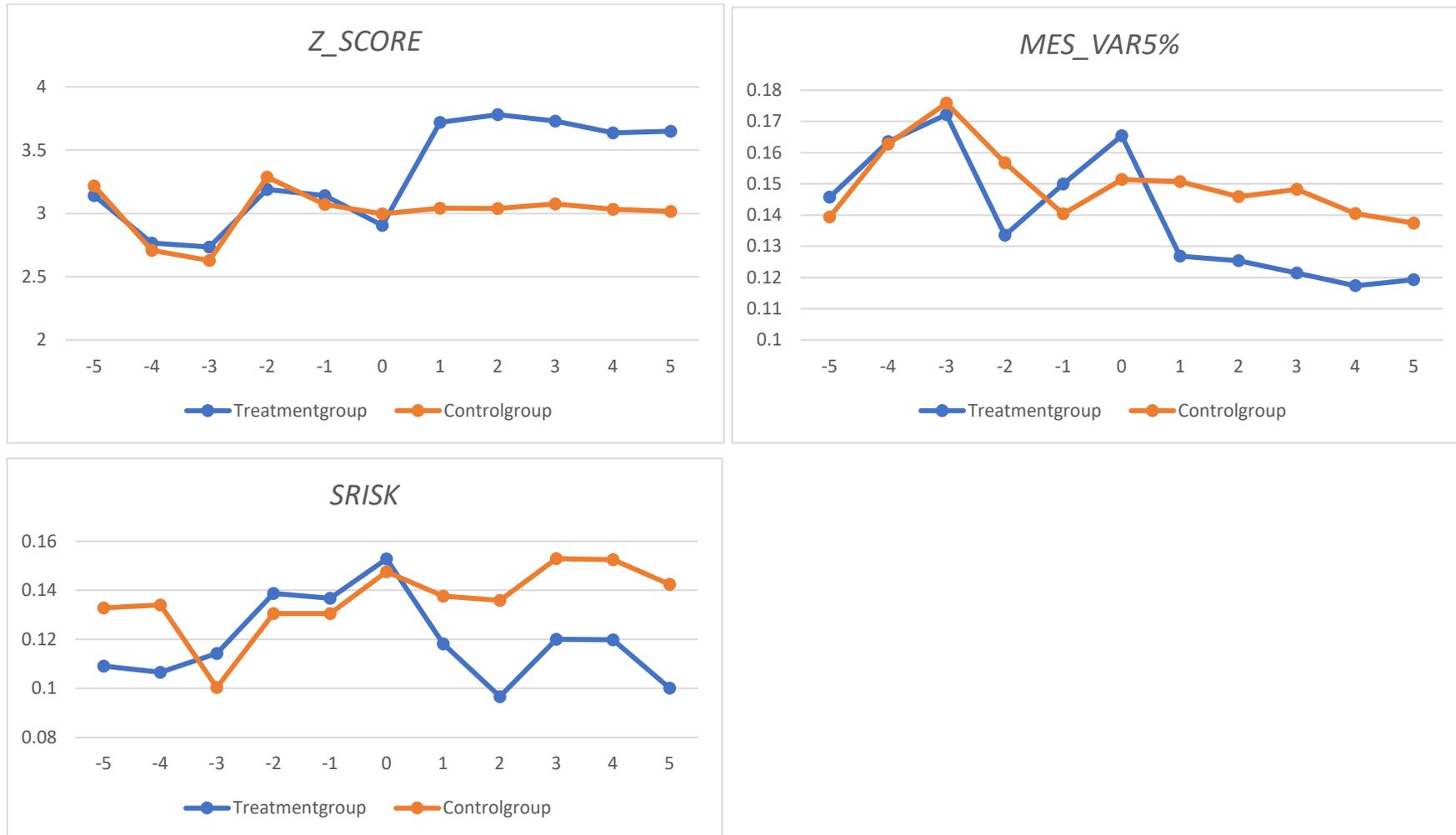

This figure shows the changes in the measures of bank stability for treatment and control groups around the enactment year of cybercrime laws. The figures are drawn from the DIDs analysis sample.

**Figure 4: Path analysis – Funding Liquidity**

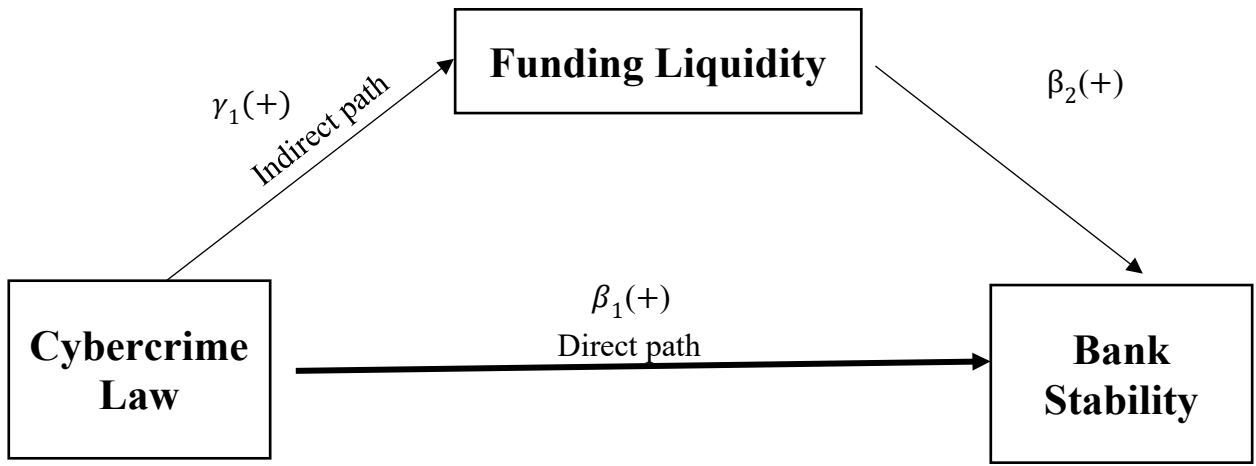

The figure shows direct and indirect paths through which cybercrime laws can affect bank stability through funding liquidity channel.

**Figure 5: Path analysis – Bank Operational Risk**

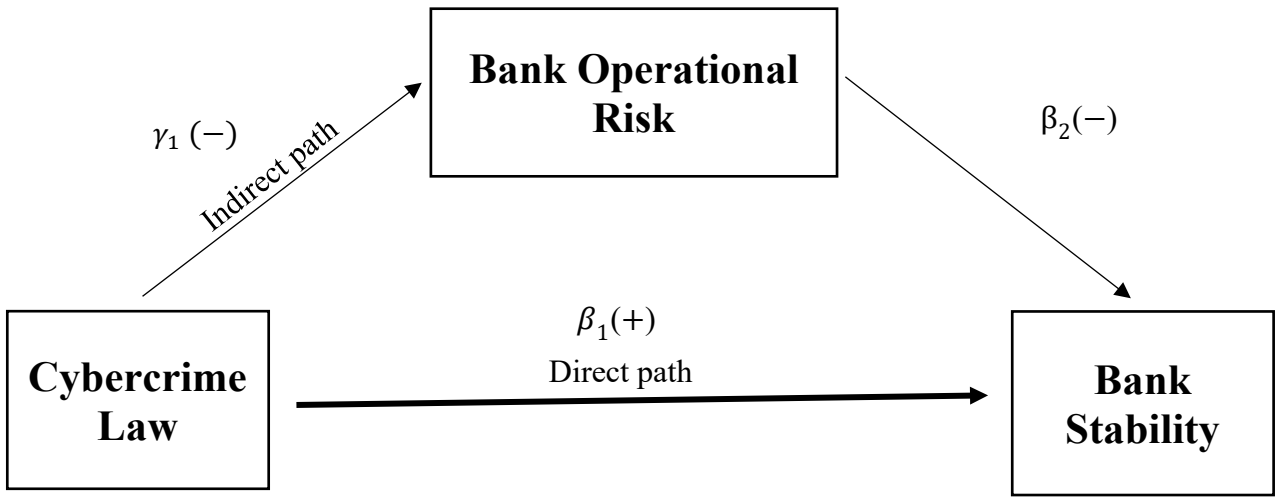

The figure illustrates the direct and indirect paths through which cybercrime laws can affect bank stability through bank operational risk channel.

**Table 1: Descriptive Statistics and Correlation Matrix**

| Panel A: Descriptive Statistics | | | | | | | | |
|---|---|---|---|---|---|---|---|---|
| Variables | N | Mean | Std. Dev | Min | P25 | Median | P75 | Max |
| Z_SCORE | 178,735 | 3.579 | 1.198 | 0.014 | 2.879 | 3.625 | 4.324 | 6.661 |
| MES_VAR5% | 6,307 | 0.125 | 0.121 | -0.042 | 0.043 | 0.107 | 0.173 | 0.933 |
| SRISK | 8,391 | 0.072 | 0.34 | 0.000 | 0.000 | 0.000 | 0.025 | 4.523 |
| CYBERLAW | 178,735 | 0.811 | 0.391 | 0.000 | 1.000 | 1.000 | 1.000 | 1.000 |
| BANK_SIZE | 178,735 | 19.728 | 1.867 | 15.808 | 18.403 | 19.411 | 20.77 | 25.433 |
| EQUITY_RATIO | 178,735 | 0.115 | 0.091 | 0.01 | 0.077 | 0.098 | 0.124 | 0.936 |
| DEPOSIT_RATIO | 178,735 | 0.853 | 0.214 | 0.004 | 0.814 | 0.936 | 0.995 | 1.000 |
| COST_EFFICIENCY | 178,735 | 0.691 | 0.245 | 0.05 | 0.573 | 0.669 | 0.772 | 2.267 |
| REV_DIV | 178,735 | 0.224 | 0.202 | -0.242 | 0.105 | 0.177 | 0.282 | 1.054 |
| EXP_DEP_INSUR | 178,735 | 0.907 | 0.29 | 0.000 | 1.000 | 1.000 | 1.000 | 1.000 |
| RULE OF LAW | 178,735 | 1.332 | 0.631 | -1.87 | 1.418 | 1.567 | 1.62 | 2.125 |
| LOG_GDP | 178,735 | 0.483 | 0.23 | -0.281 | 0.367 | 0.496 | 0.571 | 1.543 |
| LOG_GDPPERCAP | 178,735 | 0.057 | 0.101 | -0.388 | -0.001 | 0.061 | 0.098 | 0.556 |
| CREDIT_TO_GDP | 178,735 | 0.316 | 0.263 | -0.292 | 0.07 | 0.396 | 0.528 | 0.845 |

**Panel B: Correlation Matrix**

| Variables | (1) | (2) | (3) | (4) | (5) | (6) | (7) | (8) | (9) | (10) | (11) | (12) | (13) | (14) |
|---|---|---|---|---|---|---|---|---|---|---|---|---|---|---|
| *(1) Z_SCORE* | 1.000 | | | | | | | | | | | | | |
| *(2) MES_VAR5%* | -0.109 | 1.000 | | | | | | | | | | | | |
| *(3) SRISK* | -0.023 | 0.172 | 1.000 | | | | | | | | | | | |
| *(4) CYBERLAW* | 0.107 | -0.075 | -0.051 | 1.000 | | | | | | | | | | |
| *(5) BANK_SIZE* | 0.178 | 0.291 | 0.312 | 0.200 | 1.000 | | | | | | | | | |
| *(6) EQUITY_RATIO* | -0.020 | -0.037 | -0.069 | -0.007 | -0.314 | 1.000 | | | | | | | | |
| *(7) DEPOSIT_RATIO* | 0.105 | -0.110 | -0.273 | 0.131 | -0.127 | -0.242 | 1.000 | | | | | | | |
| *(8) COST_EFFICIENCY* | -0.019 | -0.025 | 0.004 | 0.127 | -0.091 | -0.136 | 0.195 | 1.000 | | | | | | |
| *(9) REV_DIV* | -0.199 | 0.092 | 0.014 | -0.244 | -0.154 | 0.177 | -0.295 | -0.167 | 1.000 | | | | | |
| *(10) EXP_DEP_INSUR* | 0.055 | -0.013 | 0.026 | 0.150 | 0.106 | 0.020 | 0.088 | 0.084 | -0.212 | 1.000 | | | | |
| *(11) RULE OF LAW* | 0.261 | -0.152 | 0.072 | 0.185 | 0.268 | -0.236 | -0.103 | 0.152 | -0.282 | 0.074 | 1.000 | | | |
| *(12) LOG_GDP* | -0.084 | -0.043 | -0.041 | 0.152 | -0.041 | 0.169 | 0.074 | -0.234 | -0.101 | 0.161 | -0.486 | 1.000 | | |
| *(13) LOG_GDPPERCAP* | -0.083 | -0.085 | -0.041 | 0.242 | -0.016 | 0.180 | 0.054 | -0.140 | -0.031 | 0.230 | -0.360 | 0.690 | 1.000 | |
| *(14) CREDIT_TO_GDP* | 0.003 | -0.141 | 0.034 | -0.026 | -0.077 | 0.022 | -0.015 | -0.127 | -0.003 | -0.068 | 0.158 | 0.033 | 0.046 | 1.000 |

This table reports the descriptive statistics and correlation matrix for all main variables used in the baseline regression analysis. Variable definitions are provided in Table A1.

| | Z_SCORE | MES_VAR5% | SRISK |
|---|---|---|---|
| | (1) | (2) | (3) |
| CYBERLAW | 0.025 | -0.026 | -0.030 |
| | [0.008] | [0.040] | [0.047] |
| BANK_SIZE | 0.002 | 0.030 | 0.049 |
| | [0.575] | [0.000] | [0.000] |
| EQUITY_RATIO | 1.348 | 0.137 | -0.169 |
| | [0.000] | [0.000] | [0.202] |
| DEPOSIT_RATIO | 0.544 | -0.025 | -0.331 |
| | [0.000] | [0.044] | [0.001] |
| COST_EFFICIENCY | -0.800 | 0.043 | 0.140 |
| | [0.000] | [0.000] | [0.023] |
| REV_DIV | -0.983 | -0.004 | -0.160 |
| | [0.000] | [0.763] | [0.102] |
| EXP_DEP_INSUR | 0.106 | 0.008 | 0.048 |
| | [0.000] | [0.415] | [0.251] |
| RULE OF LAW | 0.081 | -0.007 | 0.031 |
| | [0.031] | [0.645] | [0.438] |
| LOG_GDP | -0.576 | -0.217 | -0.013 |
| | [0.000] | [0.000] | [0.906] |
| LOG_GDPPERCAP | 0.645 | -0.030 | 0.012 |
| | [0.000] | [0.435] | [0.907] |
| CREDIT_TO_GDP | -0.136 | -0.004 | 0.089 |
| | [0.000] | [0.782] | [0.004] |
| | | | |
| Observations | 178,735 | 6,307 | 8,391 |
| Adjusted R-squared | 0.157 | 0.369 | 0.301 |
| Country FE | Yes | Yes | Yes |
| Year FE | Yes | Yes | Yes |
| Cluster by bank | Yes | Yes | Yes |

This table provides regression results of different banking system stability measures (i.e., *Z_SCORE, MES_VAR5%,* and *SRISK*) on cybercrime law enactment (*CYBERLAW*) in Columns (1)-(3), respectively. In terms of control variables, they consist of bank size (*BANK_SIZE*), the equity-to-total-assets ratio (*EQUITY_RATIO*), the deposit-to-total-assets ratio (*DEPOSIT_RATIO*), the cost-to-income ratio (*COST_EFFICIENCY*), revenue diversification (*REV_DIV*), rule of law (*RULE OF LAW*), log GDP (*LOG_GDP*), log GDP per capita (*LOG_GDPPERCAP*), and the credit-to-GDP ratio (*CREDIT_TO_GDP*). All continuous variables are winsorized at the 1st and 99th percentiles and defined in Table A1. *p-values* computed using standard errors clustered at the bank level are reported in parentheses.

| | Table 3: Difference in Differences Analysis | | |
|---|---|---|---|
| | *Z_SCORE* | *MES_VAR5%* | *SRISK* |
| | (1) | (2) | (3) |
| *POST*TREATMENT* | 0.135 | -0.045 | -0.029 |
| | [0.010] | [0.033] | [0.004] |
| *POST* | 0.046 | -0.006 | -0.004 |
| | [0.000] | [0.078] | [0.096] |
| *TREATMENT* | 0.174 | 0.035 | 0.023 |
| | [0.040] | [0.231] | [0.080] |
| *BANK_SIZE* | -0.020 | 0.029 | 0.089 |
| | [0.026] | [0.000] | [0.000] |
| *EQUITY_RATIO* | 0.962 | 0.154 | -0.043 |
| | [0.000] | [0.000] | [0.808] |
| *DEPOSIT_RATIO* | 0.548 | -0.008 | -0.208 |
| | [0.000] | [0.699] | [0.030] |
| *COST_EFFICIENCY* | -0.582 | 0.056 | 0.164 |
| | [0.000] | [0.001] | [0.041] |
| *REV_DIV* | -1.042 | -0.055 | -0.164 |
| | [0.000] | [0.074] | [0.227] |
| *EXP_DEP_INSUR* | 0.015 | 0.007 | 0.087 |
| | [0.797] | [0.631] | [0.182] |
| *RULE OF LAW* | -0.089 | 0.039 | 0.025 |
| | [0.408] | [0.358] | [0.553] |
| *LOG_GDP* | -0.465 | -0.072 | 0.255 |
| | [0.119] | [0.206] | [0.028] |
| *LOG_GDPPERCAP* | -0.124 | -0.211 | -0.266 |
| | [0.742] | [0.001] | [0.006] |
| *CREDIT_TO_GDP* | 0.023 | -0.002 | 0.133 |
| | [0.786] | [0.933] | [0.014] |
| Observations | 59,386 | 4,624 | 4,168 |
| Adjusted R-squared | 0.273 | 0.338 | 0.396 |
| Cohort*Country FE | Yes | Yes | Yes |
| Cohort*Year FE | Yes | Yes | Yes |
| Cluster by bank | Yes | Yes | Yes |

This table reports the results of the staggered difference-in-differences regressions using stacked event-by-event models suggested by Baker et al., (2022). *TREATMENT* is an indicator variable which equals 1 for countries enacting cybercrime law and zero otherwise during the sample periods. *POST* is a dummy variable equal to 1 if one to five years post law enactment and zero otherwise. Columns (1) to (3) report the results associated with the three measures of banking system stability including the Z-index (*Z_SCORE*), and the marginal expected shortfall calculated at the 5[th] percentile of value at risk (*MES_VAR5%)* and systemic risk (*SRISK*). In terms of control variables, they consist of bank size (BANK_SIZE), the equity-to-total-assets ratio *(EQUITY_RATIO*), the deposit-to-total-assets ratio (*DEPOSIT_RATIO*), the cost-to-income ratio (*COST_EFFICIENCY*), revenue diversification (*REV_DIV),* rule of law (*RULE OF LAW),* log GDP *(LOG_GDP),* log GDP per capita *(LOG_GDPPERCAP)*, and the credit-to-GDP ratio *(CREDIT_TO_GDP)*. *p-values* computed using standard errors clustered at the bank level are reported in parentheses.

| | Z_SCORE | MES_VAR5% | SRISK |
|---|---|---|---|
| **Table 4: Parallel Trend Test** | | | |
| | (1) | (2) | (3) |
| $CYBERLAW_{-5}$ | -0.048 | -0.003 | -0.004 |
| | [0.401] | [0.656] | [0.677] |
| $CYBERLAW_{-4}$ | -0.038 | 0.004 | -0.008 |
| | [0.258] | [0.563] | [0.233] |
| $CYBERLAW_{-3}$ | -0.016 | -0.014 | -0.013 |
| | [0.415] | [0.146] | [0.155] |
| $CYBERLAW_{-2}$ | -0.041 | 0.035 | 0.014 |
| | [0.170] | [0.201] | [0.104] |
| $CYBERLAW_{-1}$ | 0.083 | -0.027 | -0.016 |
| | [0.563] | [0.973] | [0.973] |
| CYBERLAW | 0.068 | -0.01 | -0.009 |
| | [0.005] | [0.055] | [0.006] |
| $CYBERLAW_{+1}$ | 0.069 | -0.004 | -0.019 |
| | [0.065] | [0.648] | [0.183] |
| $CYBERLAW_{+2}$ | 0.001 | -0.008 | -0.004 |
| | [0.997] | [0.300] | [0.684] |
| $CYBERLAW_{+3}$ | 0.026 | -0.002 | -0.005 |
| | [0.129] | [0.764] | [0.492] |
| $CYBERLAW_{+4}$ | 0.034 | 0.001 | -0.013 |
| | [0.140] | [0.931] | [0.315] |
| $CYBERLAW_{+5}$ | 0.141 | -0.001 | -0.009 |
| | [0.100] | [0.949] | [0.418] |
| BANK_SIZE | -0.001 | 0.03 | 0.048 |
| | [0.836] | [0.000] | [0.000] |
| EQUITY_RATIO | 1.34 | 0.133 | -0.16 |
| | [0.000] | [0.000] | [0.213] |
| DEPOSIT_RATIO | 0.481 | -0.005 | -0.306 |
| | [0.000] | [0.725] | [0.001] |
| COST_EFFICIENCY | -0.789 | 0.025 | 0.139 |
| | [0.000] | [0.145] | [0.022] |
| REV_DIV | -0.948 | -0.004 | -0.145 |
| | [0.000] | [0.735] | [0.118] |
| EXP_DEP_INSUR | 0.103 | -0.006 | 0.062 |
| | [0.005] | [0.606] | [0.115] |
| RULE OF LAW | 0.026 | -0.001 | 0.023 |
| | [0.679] | [0.969] | [0.559] |
| LOG_GDP | -0.6 | -0.218 | -0.007 |
| | [0.000] | [0.000] | [0.947] |
| LOG_GDPPERCAP | 0.487 | 0.014 | 0.02 |
| | [0.004] | [0.902] | [0.830] |
| CREDIT_TO_GDP | -0.094 | -0.006 | 0.086 |
| | [0.018] | [0.692] | [0.007] |
| Observations | 184,245 | 7,260 | 9,230 |
| Adjusted R-squared | 0.171 | 0.387 | 0.303 |
| Country, Year FE | Yes | Yes | Yes |
| Cluster by Bank | Yes | Yes | Yes |

This table reports the results of parallel trend tests related to the difference-in-difference regressions presented in Table 3. To validate the parallel trends assumption, we include lead and lag terms in dynamic difference-in-differences (DiD) regressions, following the methodology of Klasa et al. (2018) and Li et al. (2018). The variables $CYBERLAW_{-5}$, $CYBERLAW_{-4}$, $CYBERLAW_{-3}$, $CYBERLAW_{-2}$, and $CYBERLAW_{-1}$, indicate whether the law was enacted in the one to five years prior. The variables $CYBERLAW_{+1}$, $CYBERLAW_{+2}$, $CYBERLAW_{+3}$, $CYBERLAW_{+4}$, and $CYBERLAW_{+5}$ represent whether a country will enact the cybercrime law in one, two, three, four, or five years. Columns (1) to (3) report the results associated with the three measures of banking system stability, including the Z-index (*Z_SCORE*), and the marginal expected shortfall calculated at the 5$^{th}$ percentile of value at risk (*MES_VAR5%*) and systemic risk (*SRISK*), respectively. In terms of control variables, they consist of bank size (BANK_SIZE), the equity-to-total-assets ratio *(EQUITY_RATIO)*, the deposit-to-total-assets ratio *(DEPOSIT_RATIO)*, the cost-to-income ratio *(COST_EFFICIENCY)*, revenue diversification *(REV_DIV)*, rule of law *(RULE OF LAW)*, log GDP *(LOG_GDP)*, log GDP per capita *(LOG_GDPPERCAP)*, and the credit to GDP ratio *(CREDIT_TO_GDP)*. All continuous variables are winsorized at the 1$^{st}$ and 99$^{th}$ percentiles and defined in Table A1. *p-values* computed using standard errors clustered at the bank level are reported in parentheses.

## Table 5: Cross-sectional Analysis

**Panel A: Cybercrime Law Severity**

|  | Z_SCORE | MES_VAR5% | SRISK |
|---|---|---|---|
|  | (1) | (2) | (4) |
| CYBERLAW | 0.270 | -0.055 | 0.066 |
|  | [0.101] | [0.001] | [0.115] |
| CYBERLAW *LAW_ SEVERITY | 0.123 | -0.028 | -0.285 |
|  | [0.001] | [0.000] | [0.000] |
| Control variables | Yes | Yes | Yes |
| Observations | 178,735 | 6,307 | 8,391 |
| Adjusted R-squared | 0.152 | 0.372 | 0.306 |
| Country FE, Year FE | Yes | Yes | Yes |
| Cluster by bank | Yes | Yes | Yes |

**Panel B: Cyber Accountability**

|  | Z_SCORE | MES_VAR5% | SRISK |
|---|---|---|---|
|  | (1) | (2) | (4) |
| CYBERLAW | -0.020 | -0.013 | -0.042 |
|  | [0.164] | [0.058] | [0.195] |
| CYBERLAW * BANK_PENALTY | 0.103 | -0.035 | -0.064 |
|  | [0.000] | [0.001] | [0.087] |
| BANK_PENALTY | 0.010 | 0.005 | -0.067 |
|  | [0.707] | [0.662] | [0.070] |
| Control variables | Yes | Yes | Yes |
| Observations | 178,735 | 6,307 | 8,391 |
| Adjusted R-squared | 0.156 | 0.367 | 0.304 |
| Country FE, Year FE | Yes | Yes | Yes |
| Cluster by bank | Yes | Yes | Yes |

**Panel C: International Legal Collaboration**

|  | Z_SCORE | MES_VAR5% | SRISK |
|---|---|---|---|
|  | (1) | (2) | (4) |
| CYBERLAW | 0.051 | -0.003 | -0.027 |
|  | [0.001] | [0.685] | [0.007] |
| CYBERLAW * EXTRADITION | 0.071 | -0.015 | -0.024 |
|  | [0.000] | [0.074] | [0.034] |
| EXTRADITION | -0.043 | 0.015 | 0.017 |
|  | [0.001] | [0.063] | [0.307] |
| Control variables | Yes | Yes | Yes |
| Observations | 178,735 | 6,307 | 8,391 |
| Adjusted R-squared | 0.156 | 0.374 | 0.306 |
| Country FE, Year FE | Yes | Yes | Yes |
| Cluster by bank | Yes | Yes | Yes |

**Panel D: Law Enforcement**

|  | Z_SCORE | MES_VAR5% | SRISK |
|---|---|---|---|
|  | (1) | (2) | (4) |
| CYBERLAW | 0.080 | -0.008 | -0.036 |
|  | [0.000] | [0.340] | [0.048] |
| *CYBERLAW * RULE OF LAW* | 0.167 | -0.018 | -0.015 |
| | [0.000] | [0.003] | [0.007] |
| *RULE OF LAW* | 0.163 | -0.002 | 0.014 |
| | [0.000] | [0.854] | [0.593] |
| Control variables | Yes | Yes | Yes |
| Observations | 178,735 | 6,307 | 8,391 |
| Adjusted R-squared | 0.159 | 0.372 | 0.307 |
| Country FE, Year FE | Yes | Yes | Yes |
| Cluster by bank | Yes | Yes | Yes |

This table shows the regression results of different banking system stability measures (i.e., *Z_SCORE, MES_VAR5%,* and *SRISK*) on the interaction term between enactment of cybercrime law (*CYBERLAW*) and cybercrime law severity (*LAW_SEVERITY)* in Panel A, the interaction term between enactment of cybercrime law (*CYBERLAW*) and cyber accountability (*BANK_PENALTY)* in Panel B, the interaction term between enactment of cybercrime law (*CYBERLAW*) and international legal collaboration (*EXTRADITION)* in Panel C, and the interaction term between enactment of cybercrime law (*CYBERLAW*) and the quality of law enforcement (*RULE OF LAW)* in Panel D. In terms of control variables, they consist of bank size (BANK_SIZE), the equity-to-total-assets ratio *(EQUITY_RATIO)*, the deposit-to-total-assets ratio (*DEPOSIT_RATIO*), the cost-to-income ratio (*COST_EFFICIENCY*), revenue diversification (*REV_DIV),* rule of law (*RULE OF LAW), log GDP (LOG_GDP),* log GDP per capita *(LOG_GDPPERCAP)*, and the credit to GDP ratio *(CREDIT_TO_GDP)*. All continuous variables are winsorized at the 1st and 99th percentiles and defined in Table A1. *p-values* computed using standard errors clustered at the bank level are reported in parentheses.

## Table 6: Funding Liquidity

| | Z_SCORE | | MES_VAR5% | | SRISK | |
|---|---|---|---|---|---|---|
| **Panel B: Change in Total Funding** | | | | | | |
| | Coeff. | p-values | Coeff. | p-values | Coeff. | p-values |
| | (1) | (2) | (3) | (4) | (5) | (6) |
| **Direct Path** | | | | | | |
| P (*CYBERLAW*, Bank Stability) | 0.026 | 0.026 | -0.015 | 0.047 | -0.034 | 0.012 |
| **Indirect Path** | | | | | | |
| P (*CYBERLAW*, *ΔTOTALFUNDING*) | 0.011 | 0.000 | 0.028 | 0.001 | 0.022 | 0.038 |
| P (*ΔTOTALFUNDING*, Bank Stability) | 0.058 | 0.000 | -0.026 | 0.008 | -0.029 | 0.011 |
| P (*CYBERLAW*, *ΔTOTALFUNDING*) × P (*ΔTOTALFUNDING*, Bank Stability) | 0.001 | 0.061 | -0.001 | 0.014 | -0.001 | 0.000 |
| Total effect | 0.027 | 0.024 | -0.016 | 0.057 | -0.035 | 0.000 |
| Mediated % in Total | 2.40% | | 4.63% | | 1.84% | |
| Observations | 170,263 | | 6,160 | | 8,208 | |
| **Panel B: Change in NSFR** | Z_SCORE | | MES_VAR5% | | SRISK | |
| | Coeff. | p-values | Coeff. | p-values | Coeff. | p-values |
| | (1) | (2) | (3) | (4) | (5) | (6) |
| **Direct Path** | | | | | | |
| P (*CYBERLAW*, Bank Stability) | 0.026 | 0.000 | -0.006 | 0.018 | -0.039 | 0.005 |
| **Indirect Path** | | | | | | |
| P (*CYBERLAW*, *ΔNSFR*) | 0.020 | 0.000 | 0.014 | 0.056 | 0.017 | 0.003 |
| P (*ΔNSFR*, Bank Stability) | 0.025 | 0.000 | -0.043 | 0.014 | -0.035 | 0.021 |
| P (*CYBERLAW*, *ΔNSFR*) × P (*ΔNSFR*, Bank Stability) | 0.001 | 0.036 | -0.001 | 0.016 | -0.001 | 0.034 |
| Total effect | 0.027 | 0.032 | -0.007 | 0.018 | -0.04 | 0.005 |
| Mediated % in Total | 1.89% | | 9.12% | | 1.50% | |
| Observations | 152,139 | | 5,679 | | 7,632 | |
| **Panel C: Change in Deposits** | Z_SCORE | | MES_VAR5% | | SRISK | |
| | Coeff. | p-values | Coeff. | p-values | Coeff. | p-values |
| | (1) | (2) | (3) | (4) | (5) | (6) |
| **Direct Path** | | | | | | |

| | Coeff. | p-values | Coeff. | p-values | Coeff. | p-values |
|---|---|---|---|---|---|---|
| P (*CYBERLAW*, Bank Stability) | 0.026 | 0.030 | -0.006 | 0.022 | -0.036 | 0.010 |
| **Indirect Path** | | | | | | |
| P (*CYBERLAW*, *ΔDEPOSIT*) | 0.194 | 0.023 | 0.080 | 0.056 | 0.027 | 0.011 |
| P (*ΔDEPOSIT*, Bank Stability) | 0.025 | 0.030 | -0.005 | 0.023 | -0.035 | 0.009 |
| P (*CYBERLAW*, *ΔDEPOSIT*) × P (*ΔDEPOSIT*, Bank Stability) | 0.005 | 0.016 | -0.000 | 0.016 | -0.001 | 0.034 |
| Total effect | 0.031 | 0.032 | -0.006 | 0.018 | -0.037 | 0.005 |
| Mediated % in Total | 15.72% | | 6.25% | | 2.56% | |
| Observations | 169,783 | | 6,128 | | 8,188 | |
| **Panel D: Change in Deposit-to-Assets Ratio** | ***Z_SCORE*** | | ***MES_VAR5%*** | | ***SRISK*** | |
| | Coeff. | p-values | Coeff. | p-values | Coeff. | p-values |
| | (1) | (2) | (3) | (4) | (5) | (6) |
| **Direct Path** | | | | | | |
| P (*CYBERLAW*, Bank Stability) | 0.026 | 0.030 | -0.006 | 0.022 | -0.036 | 0.010 |
| **Indirect Path** | | | | | | |
| P (*CYBERLAW*, *ΔDEPTOASSET*) | 0.005 | 0.001 | 0.009 | 0.01 | 0.005 | 0.025 |
| P (*ΔDEPTOASSET*, Bank Stability) | 0.107 | 0.000 | -0.002 | 0.000 | -0.027 | 0.073 |
| P (*CYBERLAW*, *ΔDEPTOASSET*) × P (*ΔDEPTOASSET*, Bank Stability) | 0.001 | 0.000 | -0.000 | 0.016 | -0.000 | 0.034 |
| Total effect | 0.027 | 0.030 | -0.006 | 0.018 | -0.036 | 0.005 |
| Mediated % in Total | 2.02% | | 0.30% | | 0.37% | |
| Observations | 169,783 | | 6,128 | | 8,188 | |

This table reports path analysis results, in which we examine whether cybercrime laws can affect banking system stability through funding liquidity. We use three measures of banking system stability: *Z_SCORE*, *MES_VAR5%*, and *SRISK*. The enactment of cybercrime law is captured by *CYBERLAW*, which is a dummy variable that equals 1 if a country has enacted cybercrime laws and zero otherwise. The path is funding liquidity which is measured by four alternative bank funding liquidity metrics: (1) change in total funding (Δ*TOTALFUNDING*); (2) change in net stable funding ratio (Δ*NSFR*); (3) change in deposits (*ΔDEPOSIT*); and (4) change in deposit-to-assets ratio (*ΔDEPTOASSET*). Path analysis results associated with these four measures of funding liquidity are reported in Panels A to D, respectively. Detailed variable definitions of all other variables are provided in Table A1.

## Table 7: Bank Operational Risk

### Panel A: Bank's Overall Operational Risk

| | Z_SCORE | | MES_VAR5% | | SRISK | |
|---|---|---|---|---|---|---|
| | Coeff. | p-values | Coeff. | p-values | Coeff. | p-values |
| | (1) | (2) | (3) | (4) | (5) | (6) |
| **Direct Path** | | | | | | |
| P (*CYBERLAW*, Bank Stability) | 0.013 | 0.008 | -0.032 | 0.000 | -0.028 | 0.040 |
| **Indirect Path** | | | | | | |
| P (*CYBERLAW*, Overall Operational Risk) | -0.030 | 0.000 | -0.019 | 0.097 | -0.020 | 0.047 |
| P (Overall Operational Risk, Bank Stability) | -0.112 | 0.000 | 0.043 | 0.004 | 0.223 | 0.000 |
| P (*CYBERLAW*, Overall Operational Risk) × P (Overall Operational Risk, Bank Stability) | 0.003 | 0.026 | -0.001 | 0.000 | -0.004 | 0.068 |
| Total effect | 0.016 | 0.001 | -0.033 | 0.085 | -0.032 | 0.059 |
| Mediated % in Total | 20.54% | | 2.49% | | 13.74% | |
| Observations | 178,378 | | 6,281 | | 8,373 | |

### Panel B: Operational Risk Density

| | Z_SCORE | | MES_VAR5% | | SRISK | |
|---|---|---|---|---|---|---|
| | Coeff. | p-values | Coeff. | p-values | Coeff. | p-values |
| | (1) | (2) | (3) | (4) | (5) | (6) |
| **Direct Path** | | | | | | |
| P (*CYBERLAW*, Bank Stability) | 0.015 | 0.021 | -0.006 | 0.011 | -0.030 | 0.026 |
| **Indirect Path** | | | | | | |
| P (*CYBERLAW*, Operational Risk Density) | -0.007 | 0.071 | -0.003 | 0.066 | -0.003 | 0.076 |
| P (Operational Risk Density, Bank Stability) | -0.642 | 0.047 | 0.193 | 0.005 | 0.676 | 0.024 |
| P (*CYBERLAW*, Operational Risk Density) × P (Operational Risk Density, Bank Stability) | 0.004 | 0.011 | -0.001 | 0.039 | -0.002 | 0.036 |
| Total effect | 0.019 | 0.000 | -0.007 | 0.002 | -0.032 | 0.005 |
| Mediated % in Total | 23.05% | | 8.80% | | 6.33% | |
| Observations | 178,735 | | 6,307 | | 8,391 | |

### Panel C: Operational Risk Efficiency

|  | Z_SCORE | | MES_VAR5% | | SRISK | |
|---|---|---|---|---|---|---|
| | Coeff. | p-values | Coeff. | p-values | Coeff. | p-values |
| | (1) | (2) | (3) | (4) | (5) | (6) |
| **Direct Path** | | | | | | |
| P (*CYBERLAW*, Bank Stability) | 0.015 | 0.022 | -0.006 | 0.021 | -0.030 | 0.026 |
| **Indirect Path** | | | | | | |
| P (*CYBERLAW*, Operational Risk Efficiency) | -0.014 | 0.032 | -0.008 | 0.024 | -0.031 | 0.022 |
| P (Operational Risk Efficiency, Bank Stability) | -0.355 | 0.001 | 0.059 | 0.073 | 0.062 | 0.000 |
| P (*CYBERLAW*, Operational Risk Efficiency) × P (Operational Risk Efficiency, Bank Stability) | 0.005 | 0.000 | 0.000 | 0.002 | -0.002 | 0.021 |
| Total effect | 0.020 | 0.000 | -0.006 | 0.002 | -0.032 | 0.003 |
| Mediated % in Total | 24.89% | | 7.29% | | 6.02% | |
| Observations | 178,735 | | 6,307 | | 8,391 | |

This table reports path analysis results, in which we examine whether cybercrime laws can affect bank stability through operational risk. We use three measures of banking system stability: *Z_SCORE*, *MES_VAR5%,* and *SRISK*. The enactment of cybercrime law is captured by *CYBERLAW*, which is a dummy variable that equals 1 if a country has enacted cybercrime laws and zero otherwise. Banks' exposure to operational risk is measured by non-interest expense to total bank operational expenses (Panel A); operational risk density computed non-interest expense to total assets (Panel B); and operational risk efficiency, measured as the ratio of non-interest expense to total revenues (Panel C). Detailed variable definitions of all other variables are provided in Table A1.

| | Table 8: Effectiveness of Cybercrime Laws | |
|---|---|---|
| | *CYBER_INCIDENTS* | *INCIDENT_LOSS* |
| | (1) | (2) |
| *CYBERLAW* | -0.129 | -0.241 |
| | [0.059] | [0.045] |
| Control variables | Yes | Yes |
| Adjusted R-squared | 0.395 | 0.401 |
| Observations | 818 | 818 |
| Country FE, Year FE | Yes | Yes |
| Cluster by Country | Yes | Yes |

This table provides regression results of cyber outcomes (*CYBER_INCIDENTS* and *INDICENT_LOSS*) on the enactment cybercrime laws (*CYBERLAW*). *CYBER_INCIDENTS* is the natural logarithm of one plus the number of cyber incidents in a particular year for each country. *INCIDENT_LOSS* is the natural logarithm of one plus the total dollar loss due to cyberattacks faced by a country in a given year. *p-values* computed using standard errors cluster at the country level are reported in parentheses.

| Table A1: Variable Definition | |
|---|---|
| **Variables** | **Descriptions** |
| **Cybercrime Laws** | |
| CYBERLAW | An indicator taking the value of 1 if a country has enacted a cybercrime law and 0 otherwise. |
| LAW_SEVERITY | The natural logarithm of one plus the maximum number of years of imprisonment for a cybercrime offence. |
| BANK_PENALTY | The natural logarithm of one plus the maximum monetary penalties for non-compliance with data protection regulation. |
| CYBER_INCIDENTS | The natural logarithm of one plus the number of cyber incidents in a particular year for each country. |
| INCIDENT_LOSS | The natural logarithm of one plus the total dollar loss due to cyber-attacks faced a country in a given year. |
| **Bank Stability** | |
| Z_SCORE | Measuring how resilient a bank is to financial stress. |
| MES_VAR1% | Marginal expected shortfall calculated at the $1^{st}$ percentile of value at risk |
| MES_VAR5% | Marginal expected shortfall estimated at the $5^{th}$ percentile of value at risk. |
| SRISK | Expected capital shortfall of a bank in the face of a prolonged market downturn. |
| **Control Variables** | |
| BANK_SIZE | Natural logarithm of bank total assets. |
| EQUITY_RATIO | Equity-to-total-assets ratio. |
| DEPOSIT_RATIO | Deposit-to-total-assets ratio. |
| COST_EFFICIENCY | Ratio of total cost to total income. |
| REV_DIV | Non-interest income to total operating revenue ratio. |
| EXP_DEP_INSUR | A dummy variable which equals 1 for countries implementing deposit insurance scheme and zero otherwise. |
| RULE OF LAW | Perceptions of the extent to which people in a country are confident in and comply with the rules of society (Cope, et al., 2012). |
| LOG_GDP | Natural logarithm of GDP. |
| LOG_GDPPERCAP | Natural logarithm of GDP per capita. |
| CREDIT_TO_GDP | The credit to GDP ratio |
| **Funding Liquidity** | |
| ΔTOTALFUNDING | The change in total funding (ΔTOTALFUNDING) is calculated as the difference between the current period's total funding and the previous period's total funding, divided by the previous period's funding, and then multiplied by 100. |
| ΔNSFR | Change in net stable funding ratio (ΔNSFR) is calculated as the ratio of the percentage change in available stable funding to the percentage change in its required stable funding between two consecutive periods. The net stable funding ratio is proportion of a bank's available stable funding to its required stable funding. The available stable funding is a weighted sum of funding sources based on their stability. Required stable funding is a weighted sum of uses of funding sources based on their liquidity |
| ΔDEPOSIT | Change in deposits (ΔDEPOSIT) is computed as the difference between the current period's deposit levels and the previous period's deposit levels, expressed as a percentage change. |
| ΔDEPTOASSET | Change in the deposit-to-assets ratio (ΔDEPTOASSET) is computed as the difference between the deposit-to-assets ratio in the current period and the previous period, expressed as a percentage change. |
| VOL_DEPTOASSET | Volatility of deposit-to-assets ratio (VOL_DEPTOASSET) is estimated as the standard deviation of the deposit-to-assets ratio over a 3-year rolling window. |
| **Bank Operational Risk** | |
| Bank's Overall Operational Risk | The ratio of risk-weighted assets (RWAs) related to operational risk to total operational expenses. |
| Operational Risk Density | The ratio of operational RWAs to total assets |
| Operational Risk Efficiency | The ratio of operational RWAs to total revenues |
| **Other Variables** | |
| CYBER_INCIDENTS | The natural logarithm of one plus the number of cyber incidents in a particular year for each country. |
| INCIDENT_LOSS | The natural logarithm of one plus the total dollar loss due to cyber-attacks faced a country in a given year. |
| EXTRADITION | The natural logarithm of one plus the number of extradition treaties a country has each year. |
| HHI | Industry concentration measured by the Herfindahl-Hirschman Index. |
| FRENCH, GERMAN, SCANDINAVIAN, SOCIALIST | Countries with French civil origin, German civil origin, Scandinavian civil origin, and Socialist origin. |
| FINDEX | Level of financial development measured by the metric 'engagement in digital payments' from the World Bank, which indicates the prevalence of digital financial transactions among a country's citizens. |
| E-TRANSACTION | A dummy variable taking the value of 1 if a country has enacted e-transaction laws and 0 otherwise. |
| PROTECT_PRIVACY | A dummy variable taking the value of 1 if a country has enacted a data protection and privacy law and 0 otherwise |
| CONSUMER_PROTECT | A dummy variable taking the value of 1 if a country has enacted an online consumer protection law and 0 otherwise |
| STATE_BANK | A dummy variable which equals 1 for banks with at least 30% state ownership. |
| FOREIGN_BANK | A dummy variable which equals 1 for banks with at least 30% foreign ownership. |
| TOPSIZE | A dummy variable which equals 1 if bank size is greater than the median value of all banks' size in a country. |
| BANKDEV | The ratio of annual banking sector capitalization to GDP |
| ACTRES | Bank activity restrictions, from Barth et al. (2004), defined as the degree to which banks can participate in (a) underwriting, brokering, and trading securities, as well as all areas of the mutual fund industry, (b) insurance underwriting and sales, and (c) real estate investment, development, and management. |
| CAPREG | Bank capital regulation, from Barth et al. (2004), which represents the degree of regulatory oversight on bank capital requirements, incorporating both (1) overall capital stringency, which considers how risk factors and potential market losses are factored into minimum capital adequacy standards, and (2) initial capital stringency, which examines the types of funds allowed for a bank's initial capitalization and the requirement for their official verification. |
| SUPOWER | Supervisory power, from Barth et al. (2004), which represents the extent to which supervisory authorities have the power to implement specific measures to prevent and address issues. |

| | Table A2: Callaway and Sant'Anna (2021) Estimator | | |
|---|---|---|---|
| | Z_SCORE | MES_VAR5% | SRISK |
| | (1) | (2) | (3) |
| CYBERLAW | 0.130 | -0.037 | -0.024 |
| | [0.000] | [0.040] | [0.064] |
| Control Variables | Yes | Yes | Yes |
| Country FE | Yes | Yes | Yes |
| Year FE | Yes | Yes | Yes |
| Cluster by Bank | Yes | Yes | Yes |

This table reports estimates using the Callaway and Sant'Anna (2021) estimator. The main independent variable is the enactment of cybercrime law (*CYBERLAW*). Columns (1)-(3) report the results associated with the three measures of banking system stability, including the Z-index (*Z_SCORE*), marginal expected shortfall calculated at the 5$^{th}$ percentile of value at risk (*MES_VAR5%*), and systemic risk (*SRISK*). In terms of control variables, they consist of bank size (*BANK_SIZE*), the equity-to-total-assets ratio *(EQUITY_RATIO)*, the deposit-to-total-assets ratio (*DEPOSIT_RATIO*), the cost-to-income ratio (*COST_EFFICIENCY*), revenue diversification (*REV_DIV),* rule of law (*RULE OF LAW),* log GDP *(LOG_GDP),* log GDP per capita *(LOG_GDPPERCAP)*, and the credit-to-GDP ratio *(CREDIT_TO_GDP)*. All continuous variables are winsorized at the 1$^{st}$ and 99$^{th}$ percentiles and defined in Table A1. *p-values* computed using standard errors clustered at the bank level are reported in parentheses.

| | Z_SCORE | MES_VAR5% | SRISK |
|---|---|---|---|
| | (1) | (2) | (3) |
| *Predicted CYBERLAW* | 0.098 | -0.045 | -0.058 |
| | [0.000] | [0.000] | [0.001] |
| *BANK_SIZE* | -0.006 | 0.032 | 0.064 |
| | [0.005] | [0.000] | [0.000] |
| *EQUITY_RATIO* | 1.186 | 0.07 | -0.268 |
| | [0.000] | [0.060] | [0.053] |
| *DEPOSIT_RATIO* | 0.476 | -0.071 | -0.097 |
| | [0.000] | [0.007] | [0.654] |
| *COST_EFFICIENCY* | -0.826 | 0.036 | 0.171 |
| | [0.000] | [0.003] | [0.000] |
| *REV_DIV* | -1.001 | -0.022 | -0.066 |
| | [0.000] | [0.176] | [0.176] |
| *EXP_DEP_INSUR* | -0.18 | 0.006 | 0.215 |
| | [0.000] | [0.433] | [0.001] |
| *RULE OF LAW* | 0.156 | -0.027 | -0.133 |
| | [0.000] | [0.000] | [0.002] |
| *LOG_GDP* | -0.241 | 0.002 | 0.395 |
| | [0.027] | [0.860] | [0.000] |
| *LOG_GDPPERCAP* | 0.079 | -0.059 | -0.219 |
| | [0.541] | [0.045] | [0.045] |
| *CREDIT_TO_GDP* | 0.042 | 0.041 | 0.352 |
| | [0.184] | [0.002] | [0.002] |
| | | | |
| Observations | 152,605 | 2,893 | 5,080 |
| Adjusted R-squared | 0.187 | 0.372 | 0.359 |
| Country FE | Yes | Yes | Yes |
| Year FE | Yes | Yes | Yes |
| Cluster by Bank | Yes | Yes | Yes |
| Instrument Coefficient | 0.189 | 0.166 | 0.149 |
| [p-value] from 1st stage | [0.000] | [0.000] | [0.000] |
| Partial F–statistics for IV | 60.73 | 107.21 | 17.03 |

This table provides results of second-stage instrumented regressions using the cybercrime law enactment by the neighboring countries as an instrument for *CYBERLAW*. The dependent variables are alternative measures of bank stability, including *Z_SCORE, MES_VAR5%,* and *SRISK*. In terms of control variables, they consist of bank size (*BANK_SIZE*), the equity-to-total-assets ratio *(EQUITY_RATIO)*, the deposit-to-total-assets ratio (*DEPOSIT_RATIO*), the cost-to-income ratio (*COST_EFFICIENCY*), revenue diversification (*REV_DIV),* rule of law (*RULE OF LAW),* log GDP *(LOG_GDP),* log GDP per capita (*LOG_GDPPERCAP*), and the credit-to-GDP ratio *(CREDIT_TO_GDP)*. We report the first–stage coefficients (p-values) and F-statistics of the instrument. All continuous variables are winsorized at the 1st and 99th percentiles and defined in Table A1. *p-values* computed using standard errors clustered at the bank level are reported in parentheses.

## Table A4: Addressing Potential Confounding Effects

**Panel A: Removal of Concurrent Regulatory Reforms**

|  | Z_SCORE | MES_VAR5% | SRISK |
|---|---|---|---|
|  | (1) | (2) | (3) |
| CYBERLAW | 0.096 | -0.031 | -0.068 |
|  | [0.000] | [0.065] | [0.011] |
| Control Variables | Yes | Yes | Yes |
| Observations | 63,821 | 3,771 | 4,024 |
| Adjusted R-squared | 0.233 | 0.386 | 0.357 |
| Country FE, Year FE | Yes | Yes | Yes |
| Cluster by Bank | Yes | Yes | Yes |

**Panel B: Removal of 3-Year Window Surrounding the Regulatory Reforms**

|  | Z_SCORE | MES_VAR5% | SRISK |
|---|---|---|---|
|  | (1) | (2) | (3) |
| CYBERLAW | 0.067 | -0.035 | -0.049 |
|  | [0.048] | [0.059] | [0.095] |
| Control Variables | Yes | Yes | Yes |
| Observations | 34,732 | 2,045 | 2,258 |
| Adjusted R-squared | 0.289 | 0.385 | 0.353 |
| Country FE, Year FE | Yes | Yes | Yes |
| Cluster by Bank | Yes | Yes | Yes |

**Panel C: Removal of 5-Year Window Surrounding the Regulatory Reforms**

|  | Z_SCORE | MES_VAR5% | SRISK |
|---|---|---|---|
|  | (1) | (2) | (3) |
| CYBERLAW | 0.103 | -0.041 | -0.033 |
|  | [0.023] | [0.051] | [0.198] |
| Control Variables | Yes | Yes | Yes |
| Observations | 29,512 | 1,185 | 1,301 |
| Adjusted R-squared | 0.312 | 0.561 | 0.591 |
| Country FE, Year FE | Yes | Yes | Yes |
| Cluster by Bank | Yes | Yes | Yes |

This table provides regression results of different banking system stability measures (i.e., *Z_SCORE, MES_VAR5%,* and *SRISK*) on cybercrime law enactment (*CYBERLAW*) in Columns (1)-(4), respectively. In Panel A, we remove countries that implemented a significant banking regulation reform in the same year as cybercrime law enactment from our sample. In Panel B, we exclude countries with major banking regulation reforms occurring in the year before, the year of, or the year after cybercrime law enactment from our sample. In Panel C, we remove countries enacting significant bank regulation reforms within a five-year window of the cybercrime law enactment. In terms of control variables, they consist of bank size (*BANK_SIZE*), the equity-to-total-assets ratio (*EQUITY_RATIO*), the deposit-to-total-assets ratio (*DEPOSIT_RATIO*), the cost-to-income ratio (*COST_EFFICIENCY*), revenue diversification (*REV_DIV*), rule of law (*RULE OF LAW*), log GDP (*LOG_GDP*), log GDP per capita (*LOG_GDPPERCAP*), and the credit-to-GDP ratio (*CREDIT_TO_GDP*). All continuous variables are winsorized at the 1st and 99th percentiles and defined in Table A1. *p-values* computed using standard errors clustered at the bank level are reported in parentheses.

# Banking system stability: A global analysis of cybercrime laws

Internet Appendix

This appendix provides supplementary material to the paper. The paper refers to results reported in the "Internet Appendix" in numerous places. This appendix tabulates all such results.

| Table IA1: Country-level Summary Statistics of Bank Stability Measures |||||
| --- | --- | --- | --- | --- |
| **Country** | **N** | **Z_SCORE** | **MES_VAR5%** | **SRISK** |
| Afghanistan | 117 | 2.474 | | |
| Albania | 282 | 2.087 | | |
| Algeria | 194 | 2.709 | | |
| Angola | 142 | 3.523 | 0.097 | 0.004 |
| Argentina | 1,822 | 2.020 | 0.317 | 0.075 |
| Armenia | 263 | 2.888 | | |
| Australia | 1,279 | 3.706 | 0.103 | 0.371 |
| Austria | 7,540 | 3.804 | 0.120 | 0.064 |
| Azerbaijan | 235 | 2.619 | | |
| Bahrain | 102 | 2.609 | | |
| Bangladesh | 99 | 2.510 | 0.250 | 0.000 |
| Belize | 128 | 2.349 | | |
| Benin | 804 | 2.788 | 0.168 | 0.011 |
| Bhutan | 432 | 2.658 | 0.199 | 0.000 |
| Bosnia and Herzegovina | 195 | 3.422 | 0.124 | 0.001 |
| Botswana | 474 | 3.244 | | 0.000 |
| Brazil | 41 | 3.116 | | |
| Brunei Darussalam | 2,152 | 2.514 | 0.174 | 0.166 |
| Bulgaria | 61 | 3.250 | | |
| Burkina Faso | 61 | 3.159 | | |
| Burundi | 222 | 2.799 | 0.137 | 0.001 |
| Cameroon | 290 | 3.608 | | 0.869 |
| Canada | 9,317 | 4.194 | 0.047 | 0.012 |
| Chile | 323 | 2.907 | 0.100 | 0.013 |
| China | 1,845 | 3.499 | 0.133 | 0.520 |
| Colombia | 231 | 2.433 | 0.287 | 0.000 |
| Costa Rica | 149 | 2.545 | | |
| Croatia | 823 | 2.916 | 0.186 | 0.000 |
| Cuba | 907 | 3.385 | | |
| Cyprus | 51 | 2.797 | | |
| Czech Republic | 225 | 2.381 | 0.301 | 0.149 |
| Côte D'Ivoire | 674 | 2.891 | 0.228 | 0.010 |
| Denmark | 16,891 | 4.141 | 0.127 | 1.736 |
| Djibouti | 53 | 2.862 | | |
| Dominican Republic | 1,816 | 3.168 | 0.062 | 0.113 |
| Ecuador | 646 | 2.924 | | |
| Egypt | 318 | 3.119 | | |
| El Salvador | 438 | 2.986 | | |
| Estonia | 912 | 2.706 | 0.144 | 0.008 |
| Ethiopia | 3,619 | 3.426 | 0.194 | 0.777 |
| Finland | 128 | 2.554 | 0.137 | 0.002 |
| France | 230 | 3.244 | | |
| Gambia | 364 | 3.764 | 0.159 | 0.016 |
| Germany | 6,465 | 3.176 | 0.135 | 1.941 |
| Ghana | 5,593 | 2.942 | 0.104 | 1.257 |

| Country | | | | |
|---|---:|---:|---:|---:|
| Greece | 472 | 2.491 | 0.182 | 0.000 |
| Guatemala | 43 | 2.362 | | |
| Guinea | 67 | 2.259 | | 0.000 |
| Honduras | 402 | 2.466 | 0.376 | 0.163 |
| Hungary | 484 | 3.069 | | |
| Iceland | 322 | 3.444 | | |
| India | 843 | 2.781 | 0.136 | 0.001 |
| Indonesia | 952 | 2.616 | 0.209 | 0.015 |
| Iran (Islamic Republic Of) | 1,280 | 3.258 | 0.115 | 0.014 |
| Ireland | 2,442 | 3.105 | 0.200 | 0.039 |
| Israel | 491 | 2.931 | | |
| Italy | 279 | 2.603 | | |
| Jamaica | 143 | 2.256 | 0.227 | 0.000 |
| Japan | 258 | 3.670 | 0.146 | 0.090 |
| Jordan | 13,770 | 3.391 | 0.150 | 0.596 |
| Kazakhstan | 204 | 3.073 | | |
| Kenya | 99 | 3.313 | 0.089 | 0.002 |
| Kuwait | 13,752 | 3.659 | 0.130 | 0.129 |
| Kyrgyzstan | 307 | 2.802 | 0.204 | 0.004 |
| Lao People's Democratic Republic | 809 | 2.924 | 0.107 | 0.001 |
| Latvia | 178 | 2.930 | | |
| Lebanon | 1,282 | 2.928 | 0.135 | 0.111 |
| Lesotho | 122 | 3.447 | 0.110 | 0.002 |
| Liechtenstein | 82 | 2.961 | | |
| Lithuania | 522 | 2.903 | 0.157 | 0.010 |
| Luxembourg | 212 | 3.289 | 0.138 | 0.003 |
| Madagascar | 506 | 3.169 | 0.154 | 0.008 |
| Malawi | 60 | 2.699 | | |
| Malaysia | 187 | 2.481 | 0.370 | 0.000 |
| Mali | 2,564 | 3.112 | 0.265 | |
| Malta | 406 | 2.123 | | |
| Mauritania | 252 | 2.825 | | |
| Mauritius | 92 | 2.669 | | |
| Mexico | 771 | 2.700 | 0.303 | 0.027 |
| Mongolia | 319 | 2.850 | | 0.001 |
| Montenegro | 137 | 2.609 | | |
| Mozambique | 269 | 3.361 | 0.076 | 0.067 |
| Nepal | 177 | 2.767 | | |
| Netherlands | 126 | 3.036 | | |
| New Zealand | 249 | 2.206 | | |
| Nicaragua | 174 | 3.332 | | |
| Niger | 314 | 2.886 | 0.057 | 0.000 |
| Nigeria | 169 | 2.355 | 0.134 | 0.000 |
| Norway | 1,492 | 3.277 | 0.122 | 0.028 |
| Oman | 64 | 2.759 | | |
| Panama | 748 | 2.611 | 0.125 | 0.004 |
| Paraguay | 193 | 2.758 | | |
| Peru | 576 | 3.447 | 0.157 | 2.564 |

| Country | | | | |
|---|---:|---:|---:|---:|
| Philippines | 2,118 | 3.697 | 0.072 | 0.026 |
| Poland | 796 | 3.298 | | |
| Portugal | 213 | 3.744 | 0.749 | |
| Republic Of Korea | 247 | 3.343 | 0.101 | 0.000 |
| Republic Of Moldova | 536 | 2.944 | 0.159 | 0.002 |
| Romania | 477 | 3.046 | 0.150 | 0.012 |
| Rwanda | 782 | 3.329 | 0.098 | 0.009 |
| Saudi Arabia | 1,506 | 3.408 | 0.170 | 0.005 |
| Senegal | 1,310 | 3.153 | 0.213 | 0.159 |
| Serbia | 328 | 2.906 | | |
| Seychelles | 462 | 2.287 | 0.216 | 0.000 |
| Sierra Leone | 134 | 2.551 | | 0.013 |
| Singapore | 61 | 3.565 | 0.149 | 0.001 |
| Slovakia | 279 | 2.809 | | |
| Slovenia | 199 | 2.665 | 0.278 | |
| South Africa | 460 | 3.437 | 0.097 | 0.009 |
| Spain | 104 | 2.254 | | |
| Sri Lanka | 201 | 3.688 | | |
| Sudan | 554 | 2.386 | 0.259 | 0.007 |
| Sweden | 303 | 2.843 | 0.286 | 0.005 |
| Switzerland | 343 | 3.000 | 0.412 | 0.027 |
| Tajikistan | 1,505 | 3.643 | 0.189 | 0.799 |
| Thailand | 95 | 2.501 | | |
| Republic Of Macedonia | 85 | 2.589 | 0.446 | 0.001 |
| Togo | 759 | 2.875 | 0.172 | 0.026 |
| Trinidad And Tobago | 65 | 2.669 | | |
| Turkey | 135 | 3.309 | 0.083 | 0.000 |
| Uganda | 658 | 2.644 | 0.236 | 0.003 |
| Ukraine | 520 | 2.608 | 0.107 | 0.000 |
| United Arab Emirates | 401 | 2.587 | 0.076 | 0.000 |
| United Kingdom | 964 | 2.461 | 0.192 | 0.000 |
| United Republic of Tanzania | 460 | 2.177 | | |
| United States of America | 105,524 | 3.634 | | 0.043 |
| Uruguay | 124 | 3.110 | | |
| Uzbekistan | 606 | 2.449 | 0.478 | |
| Bolivarian Republic of Venezuela | 535 | 3.452 | 0.164 | 0.015 |
| Viet Nam | 632 | 2.897 | 0.142 | 0.067 |
| Zambia | 187 | 2.081 | 0.264 | 0.003 |
| Zimbabwe | 151 | 1.826 | | |

This table reports country-level descriptive statistics for all bank stability measures used in this study. Detailed definitions of these variables are provided in Table A1.

## Table IA2: Different Fixed Effect Models

### Panel A: Controlling for Bank Fixed Effects

|  | Z_SCORE | MES_VAR5% | SRISK |
|---|---|---|---|
|  | (1) | (2) | (3) |
| CYBERLAW | 0.024 | -0.017 | -0.025 |
|  | [0.020] | [0.011] | [0.007] |
| Control Variables | Yes | Yes | Yes |
| Observations | 178,735 | 6,307 | 8,391 |
| Adjusted R-squared | 0.492 | 0.281 | 0.437 |
| Bank FE | Yes | Yes | Yes |
| Year FE | Yes | Yes | Yes |
| Cluster by Bank | Yes | Yes | Yes |

### Panel B: Controlling for Bank*Country Fixed Effects

|  | Z_SCORE | MES_VAR5% | SRISK |
|---|---|---|---|
|  | (1) | (2) | (3) |
| CYBERLAW | 0.019 | -0.027 | -0.010 |
|  | [0.085] | [0.002] | [0.003] |
| Control Variables | Yes | Yes | Yes |
| Observations | 178,735 | 6,307 | 8,391 |
| Adjusted R-squared | 0.486 | 0.295 | 0.471 |
| Bank*Country FE | Yes | Yes | Yes |
| Year FE | Yes | Yes | Yes |
| Cluster by Bank | Yes | Yes | Yes |

This table provides regression results of different banking system stability measures (i.e., *Z_SCORE, MES_VAR5%,* and *SRISK*) on cybercrime law enactment (*CYBERLAW*) in Columns (1)-(3), respectively. We account for bank fixed effects and bank*country fixed effects in Panels A and B, respectively. In terms of control variables, they consist of bank size (*BANK_SIZE*), the equity-to-total-assets ratio *(EQUITY_RATIO)*, the deposit-to-total-assets ratio (*DEPOSIT_RATIO*), the cost-to-income ratio (*COST_EFFICIENCY*), revenue diversification (*REV_DIV)*, rule of law (*RULE OF LAW),* log GDP *(LOG_GDP),* log GDP per capita *(LOG_GDPPERCAP)*, and the credit-to-GDP ratio *(CREDIT_TO_GDP)*. All continuous variables are winsorized at the 1$^{st}$ and 99$^{th}$ percentiles and defined in Table A1. *p-values* computed using standard errors clustered at the bank level are reported in parentheses.

# Table IA3: Volatility of Deposit-to-Assets Ratio

|  | Z_SCORE | | MES_VAR5% | | SRISK | |
|---|---|---|---|---|---|---|
|  | Coeff. | p-values | Coeff. | p-values | Coeff. | p-values |
|  | (1) | (2) | (3) | (4) | (5) | (6) |
| **Direct Path** | | | | | | |
| P (CYBERLAW, Bank Stability) | 0.033 | 0.009 | -0.009 | 0.063 | -0.033 | 0.019 |
| **Indirect Path** | | | | | | |
| P (CYBERLAW, VOL_DEPTOASSET) | -0.004 | 0.000 | -0.008 | 0.022 | -0.007 | 0.008 |
| P (VOL_DEPTOASSET, Bank Stability) | -1.415 | 0.000 | 0.009 | 0.069 | 0.401 | 0.000 |
| P (CYBERLAW, VOL_DEPTOASSET) × P (VOL_DEPTOASSET, Bank Stability) | 0.006 | 0.026 | -0.000 | 0.069 | -0.003 | 0.034 |
| Total Effect | 0.039 | 0.009 | -0.009 | 0.063 | -0.036 | 0.005 |
| Mediated % in Total | 14.64% | | 0.79% | | 7.84% | |
| Observations | 156,498 | | 5,390 | | 7,379 | |

This table reports path analysis results, in which we examine whether cybercrime laws can affect banking system stability through funding liquidity. We measure funding liquidity as the standard deviation of the deposit-to-assets ratio over a rolling window (i.e., volatility of deposit-to-assets ratio), which provides insight into the stability of the deposit funding source. The lower the volatility of the deposit-to-asset ratio (*VOL_DEPTOASSET*), the more stable the funding source. We use three measures of banking system stability: *Z_SCORE*, *MES_VAR5%,* and *SRISK*. Detailed variable definitions of all other variables are provided in Table A1.

| | Table IA4: Country Legal Origin | | |
|---|---|---|---|
| | Z_SCORE | MES_VAR5% | SRISK |
| | (1) | (2) | (3) |
| CYBERLAW | 0.039 | -0.040 | -0.123 |
| | [0.405] | [0.000] | [0.001] |
| CYBERLAW*FRENCH | -0.041 | 0.010 | 0.030 |
| | [0.438] | [0.578] | [0.562] |
| CYBERLAW*GERMAN | -0.277 | 0.056 | 0.118 |
| | [0.000] | [0.005] | [0.005] |
| CYBERLAW* SCANDINAVIAN | 0.282 | 0.045 | -0.016 |
| | [0.000] | [0.008] | [0.797] |
| CYBERLAW* SOCIALIST | 0.242 | -0.003 | -0.189 |
| | [0.001] | [0.905] | [0.000] |
| Control Variables | Yes | Yes | Yes |
| Observations | 178,735 | 6,307 | 8,391 |
| Adjusted R-squared | 0.127 | 0.284 | 0.248 |
| Country FE, Year FE | Yes | Yes | Yes |
| Cluster by Bank | Yes | Yes | Yes |

This table provides regression results of different banking system stability measures (i.e., *Z_SCORE, MES_VAR5%,* and *SRISK*) on the interaction term between enactment of cybercrime law (*CYBERLAW*) and the legal origin (*FRENCH, GERMAN, SCANDINAVIAN, SOCIALIST)*. Using English common law as the benchmark, we omit it from our models. All continuous variables are winsorized at the 1st and 99th percentiles and defined in Table A1. *p-values* computed using standard errors cluster at the bank level are reported in parentheses.

|  | Z_SCORE | MES_VAR5% | SRISK |
|---|---|---|---|
|  | (1) | (2) | (3) |
| *CYBERLAW* | -0.280 | -0.051 | -0.014 |
|  | [0.000] | [0.133] | [0.918] |
| *CYBERLAW*FINDEX* | 0.212 | -0.027 | 0.119 |
|  | [0.000] | [0.090] | [0.763] |
| *FINDEX* | -4.035 | -0.061 | -0.226 |
|  | [0.001] | [0.457] | [0.531] |
| Control Variables | Yes | Yes | Yes |
| Observations | 21,310 | 657 | 924 |
| Adjusted R-squared | 0.106 | 0.519 | 0.148 |
| Country FE, Year FE | Yes | Yes | Yes |
| Cluster by Bank | Yes | Yes | Yes |

**Table IA5: Fintech Development**

This table provides regression results of different banking system stability measures (i.e., *Z_SCORE, MES_VAR5%,* and *SRISK*) on the interaction term between enactment of cybercrime law (*CYBERLAW*) and the financial development index (*FINDEX*). All continuous variables are winsorized at the 1st and 99th percentiles and defined in Table A1. *p-values* computed using standard errors clustered at the bank level are reported in parentheses.

| | Table IA6: Moral Hazard | | |
|---|---|---|---|
| | Z_SCORE | MES_VAR5% | SRISK |
| | (1) | (2) | (3) |
| CYBERLAW | 0.054 | -0.314 | -0.039 |
| | [0.073] | [0.137] | [0.136] |
| CYBERLAW* EXP_DEP_INSUR | 0.024 | 0.252 | -0.143 |
| | [0.487] | [0.198] | [0.101] |
| EXP_DEP_INSUR | -0.041 | -0.097 | 0.150 |
| | [0.194] | [0.497] | [0.084] |
| Control Variables | Yes | Yes | Yes |
| Observations | 178,735 | 6,307 | 8,391 |
| Adjusted R-squared | 0.152 | 0.361 | 0.295 |
| Country FE, Year FE | Yes | Yes | Yes |
| Cluster by Bank | Yes | Yes | Yes |

This table provides regression results of different banking system stability measures (i.e., *Z_SCORE, MES_VAR5%,* and *SRISK*) on the interaction term between enactment of cybercrime law (*CYBERLAW*) and a dummy variable indicating if a country has an explicit deposit insurance system (*EXP_DEP_INSUR*). *EXP_DEP_INSUR* equals 1 if a country has an explicit deposit insurance system and 0 otherwise. All continuous variables are winsorized at the 1st and 99th percentiles and defined in Table A1. *p-values* computed using standard errors clustered at the bank level are reported in parentheses.

## Table IA7: Cyber Law Heterogeneity

**Panel A: E-Transaction Laws**

|  | Z_SCORE | MES_VAR5% | SRISK |
|---|---|---|---|
|  | (1) | (2) | (3) |
| CYBERLAW | 0.067 | -0.021 | -0.03 |
|  | [0.029] | [0.028] | [0.008] |
| CYBERLAW*E_TRANSACTION | -0.093 | 0.017 | 0.001 |
|  | [0.003] | [0.367] | [0.964] |
| E_TRANSACTION | -0.004 | -0.002 | -0.004 |
|  | [0.868] | [0.845] | [0.853] |
| Control Variables | Yes | Yes | Yes |
| Observations | 178,735 | 6,307 | 8,391 |
| Adjusted R-squared | 0.157 | 0.367 | 0.300 |
| Country FE; Year FE | Yes | Yes | Yes |
| Cluster by Bank | Yes | Yes | Yes |

**Panel B: Data Protection and Privacy Laws**

|  | Z_SCORE | MES_VAR5% | SRISK |
|---|---|---|---|
|  | (1) | (2) | (3) |
| CYBERLAW | 0.053 | -0.003 | -0.042 |
|  | [0.001] | [0.718] | [0.037] |
| CYBERLAW*PROTECT_PRIVACY | -0.133 | -0.002 | -0.055 |
|  | [0.000] | [0.887] | [0.166] |
| PROTECT_PRIVACY | -0.088 | 0.028 | -0.059 |
|  | [0.006] | [0.056] | [0.099] |
| Control Variables | Yes | Yes | Yes |
| Observations | 178,735 | 6,307 | 8,391 |
| Adjusted R-squared | 0.157 | 0.365 | 0.298 |
| Country FE; Year FE | Yes | Yes | Yes |
| Cluster by Bank | Yes | Yes | Yes |

**Panel C: Consumer Protection Regulations**

|  | Z_SCORE | MES_VAR5% | SRISK |
|---|---|---|---|
|  | (1) | (2) | (3) |
| CYBERLAW | 0.029 | -0.001 | -0.044 |
|  | [0.109] | [0.029] | [0.051] |
| CYBERLAW *CONSUMER_PROTECT | -0.090 | -0.009 | 0.028 |
|  | [0.000] | [0.414] | [0.221] |
| CONSUMER_PROTECT | 0.011 | 0.009 | -0.031 |
|  | [0.627] | [0.327] | [0.157] |
| Control Variables | Yes | Yes | Yes |
| Observations | 178,735 | 6,307 | 8,391 |
| Adjusted R-squared | 0.155 | 0.366 | 0.299 |
| Country FE; Year FE | Yes | Yes | Yes |
| Cluster by Bank | Yes | Yes | Yes |

This table shows regression results of different banking system stability measures (i.e., Z_SCORE, MES_VAR5%, and SRISK) on the interaction term between enactment of cybercrime law (CYBERLAW) and E-Transaction law (E_TRANSACTION) in Panel A, the interaction term between enactment of cybercrime law (CYBERLAW) and data protection and privacy laws (PROTECT_PRIVACY) in Panel B, and the interaction term between enactment of cybercrime law (CYBERLAW) and consumer protection regulations (CONSUMER_PROTECT) in Panel C. In terms of control variables, they consist of bank size (BANK_SIZE), the equity-to-total-assets ratio (EQUITY_RATIO), the deposit-to-total-assets ratio (DEPOSIT_RATIO), the cost-to-income ratio (COST_EFFICIENCY), revenue diversification (REV_DIV), rule of law (RULE OF LAW), log GDP (LOG_GDP), log GDP per capita (LOG_GDPPERCAP), and the credit-to-GDP ratio (CREDIT_TO_GDP). All continuous variables are winsorized at the 1st and 99th percentiles and defined in Table A1. *p-values* computed using standard errors clustered at the bank level are reported in parentheses.

# Table IA8: Banking Industry Concentration

|  | Z_SCORE | MES_VAR5% | SRISK |
|---|---|---|---|
|  | (1) | (2) | (4) |
| *CYBERLAW* | -0.062 | 0.003 | -0.032 |
|  | [0.002] | [0.008] | [0.097] |
| *CYBERLAW*HHI* | 0.540 | -0.090 | -0.017 |
|  | [0.002] | [0.046] | [0.854] |
| *HHI* | -0.324 | 0.145 | -0.040 |
|  | [0.071] | [0.076] | [0.795] |
| Control Variables | Yes | Yes | Yes |
| Observations | 178,735 | 6,307 | 8,391 |
| Adjusted R-squared | 0.157 | 0.370 | 0.301 |
| Country FE, Year FE | Yes | Yes | Yes |
| Cluster by Bank | Yes | Yes | Yes |

This table shows regression results of different banking system stability measures (i.e., Z_SCORE, MES_VAR5%, and SRISK) on the interaction term between the enactment of cybercrime law (*CYBERLAW*) and industry concentration (*HHI*). *HHI* refers to the Herfindahl–Hirschman Index, used to measure the level of industry concentration. A higher index is associated with higher industry concentration. In terms of control variables, they consist of bank size (BANK_SIZE), the equity-to-total-assets ratio *(EQUITY_RATIO)*, the deposit-to-total-assets ratio (*DEPOSIT_RATIO*), the cost-to-income ratio (*COST_EFFICIENCY*), revenue diversification (*REV_DIV),* rule of law (*RULE OF LAW),* log GDP *(LOG_GDP),* log GDP per capita *(LOG_GDPPERCAP)*, and the credit-to-GDP ratio *(CREDIT_TO_GDP)*. All continuous variables are winsorized at the 1st and 99th percentiles and defined in Table A1. *p-values* computed using standard errors clustered at the bank level are reported in parentheses.

| | Table IA9: State Ownership | | |
|---|---|---|---|
| | Z_SCORE | MES_VAR5% | SRISK |
| | (1) | (2) | (3) |
| CYBERLAW | -0.017 | -0.003 | -0.055 |
| | [0.128] | [0.641] | [0.003] |
| CYBERLAW*STATE_BANK | 0.147 | -0.053 | 0.168 |
| | [0.070] | [0.012] | [0.201] |
| STATE_BANK | 0.056 | -0.031 | -0.026 |
| | [0.394] | [0.087] | [0.035] |
| Control Variables | Yes | Yes | Yes |
| Observations | 178,735 | 6,307 | 8,391 |
| Adjusted R-squared | 0.154 | 0.370 | 0.303 |
| Country FE, Year FE | Yes | Yes | Yes |
| Cluster by Bank | Yes | Yes | Yes |

This table provides regression results of different banking system stability measures (i.e., Z_SCORE, MES_VAR5%, and SRISK) on the interaction term between enactment of cybercrime law (CYBERLAW) and bank state ownership (STATE_BANK). STATE_BANK is a dummy variable taking the value of 1 if at least 30% of bank shares are owned by a government and 0 otherwise. All continuous variables are winsorized at the 1st and 99th percentiles and defined in Table A1. *p-values* computed using standard errors clustered at the bank level are reported in parentheses.

| | Table IA10: Foreign Ownership | | |
|---|---|---|---|
| | Z_SCORE | MES_VAR5% | SRISK |
| | (1) | (2) | (3) |
| *CYBERLAW* | 0.028 | -0.001 | -0.052 |
| | [0.043] | [0.882] | [0.002] |
| *CYBERLAW*FOREIGN_BANK* | 0.014 | -0.054 | -0.047 |
| | [0.839] | [0.004] | [0.089] |
| *FOREIGN_BANK* | -0.040 | 0.034 | 0.020 |
| | [0.525] | [0.021] | [0.092] |
| Control Variables | Yes | Yes | Yes |
| Observations | 178,735 | 6,307 | 8,391 |
| Adjusted R-squared | 0.156 | 0.372 | 0.301 |
| Country FE, Year FE | Yes | Yes | Yes |
| Cluster by Bank | Yes | Yes | Yes |

This table provides regression results of different banking system stability measures (i.e., *Z_SCORE, MES_VAR5%,* and *SRISK*) on the interaction term between enactment of cybercrime law (*CYBERLAW*) and bank foreign ownership (*FOREIGN_BANK*). *FOREIGN_BANK* is a dummy variable taking the value of 1 if at least 30% of bank shares are owned by a foreign entity and 0 otherwise. All continuous variables are winsorized at the 1st and 99th percentiles and defined in Table A1. *p-values* computed using standard errors clustered at the bank level are reported in parentheses.

| | Z_SCORE | MES_VAR5% | SRISK | Z_SCORE | MES_VAR5% | SRISK | Z_SCORE | MES_VAR5% | SRISK |
|---|---|---|---|---|---|---|---|---|---|
| | (1) | (2) | (3) | (4) | (5) | (6) | (7) | (8) | (9) |
| CYBERLAW*CAPREG | -0.024 | 0.004 | 0.034 | | | | | | |
| | [0.011] | [0.111] | [0.000] | | | | | | |
| CAPREG | 0.031 | -0.013 | -0.012 | | | | | | |
| | [0.001] | [0.000] | [0.098] | | | | | | |
| CYBERLAW*ACTRES | | | | -0.026 | 0.003 | 0.023 | | | |
| | | | | [0.000] | [0.296] | [0.081] | | | |
| ACTRES | | | | 0.01 | -0.005 | -0.023 | | | |
| | | | | [0.144] | [0.037] | [0.118] | | | |
| CYBERLAW*SUPOWER | | | | | | | -0.035 | 0.002 | 0.040 |
| | | | | | | | [0.000] | [0.540] | [0.091] |
| SUPOWER | | | | | | | -0.003 | -0.006 | -0.045 |
| | | | | | | | [0.616] | [0.075] | [0.018] |
| CYBERLAW | 0.183 | -0.068 | -0.118 | 0.137 | -0.009 | -0.248 | 0.562 | -0.052 | -0.514 |
| | [0.001] | [0.000] | [-0.034] | [0.004] | [0.689] | [0.031] | [0.000] | [0.197] | [0.056] |
| Control Variables | Yes | Yes | Yes | Yes | Yes | Yes | Yes | Yes | Yes |
| Observations | 161,698 | 5,608 | 7,316 | 161,698 | 5,608 | 7,316 | 161,698 | 5,608 | 7,316 |
| Adjusted R-squared | 0.153 | 0.380 | 0.306 | 0.154 | 0.384 | 0.307 | 0.154 | 0.379 | 0.308 |
| Country FE, Year FE | Yes | Yes | Yes | Yes | Yes | Yes | Yes | Yes | Yes |
| Cluster by Bank | Yes | Yes | Yes | Yes | Yes | Yes | Yes | Yes | Yes |

Table IA11: Banking Regulations

This table provides regression results of different banking system stability measures (i.e., Z_SCORE, MES_VAR5%, and SRISK) on the interaction term between enactment of cybercrime law (CYBERLAW) and bank capital regulation (CAPREG) from Columns (1)-(3), the interaction term between enactment of cybercrime law (CYBERLAW) and bank activity restrictions (ACTRES) from Columns (4)-(6), and the interaction term between enactment of cybercrime law (CYBERLAW) and bank supervisory power (SUPOWER) from Columns (7)-(9). All continuous variables are winsorized at the 1st and 99th percentiles and defined in Table A1. *p-values* computed using standard errors clustered at the bank level are reported in parentheses.

| | Table IA12: Bank Size | | |
|---|---|---|---|
| | Z_SCORE | MES_VAR5% | SRISK |
| | (1) | (2) | (3) |
| CYBERLAW | 0.059 | -0.011 | 0.011 |
| | [0.002] | [0.357] | [0.577] |
| CYBERLAW*TOPSIZE | 0.083 | -0.001 | -0.052 |
| | [0.000] | [0.124] | [0.020] |
| TOPSIZE | -0.049 | 0.037 | 0.055 |
| | [0.020] | [0.000] | [0.002] |
| Control Variables | Yes | Yes | Yes |
| Observations | 178,735 | 6,307 | 8,391 |
| Adjusted R-squared | 0.157 | 0.367 | 0.297 |
| Country FE, Year FE | Yes | Yes | Yes |
| Cluster by Bank | Yes | Yes | Yes |

This table provides regression results of different banking system stability measures (i.e., *Z_SCORE, MES_VAR5%,* and *SRISK*) on the interaction term between enactment of cybercrime law (*CYBERLAW*) and bank size (*TOPSIZE*). *TOPSIZE* is a dummy variable which equals 1 if the bank size is greater than the median value of all banks' sizes in a country and 0 otherwise. All continuous variables are winsorized at the 1$^{st}$ and 99$^{th}$ percentiles and defined in Table A1. *p-values* computed using standard errors clustered at the bank level are reported in parentheses.

|  | Table IA13: Banking Sector Development | | |
|---|---|---|---|
|  | Z_SCORE | MES_VAR5% | SRISK |
|  | (1) | (2) | (3) |
| CYBERLAW | -0.007 | -0.017 | -0.019 |
|  | [0.501] | [0.030] | [0.149] |
| CYBERLAW*BANKDEV | 0.129 | -0.044 | -0.041 |
|  | [0.003] | [0.021] | [0.026] |
| BANKDEV | -0.224 | -0.001 | 0.151 |
|  | [0.000] | [0.955] | [0.000] |
| Control Variables | Yes | Yes | Yes |
| Observations | 178,735 | 6,307 | 8,391 |
| Adjusted R-squared | 0.157 | 0.369 | 0.301 |
| Country FE, Year FE | Yes | Yes | Yes |
| Cluster by Bank | Yes | Yes | Yes |

This table provides regression results of different banking system stability measures (i.e., *Z_SCORE, MES_VAR5%,* and *SRISK*) on the interaction term between enactment of cybercrime law (*CYBERLAW*) and the banking sector development (*BANKDEV*). *BANKDEV* is the ratio of annual banking sector capitalization to GDP, which reflects the banking sector development in each country (Delis and Tsionas, 2009). All continuous variables are winsorized at the 1$^{st}$ and 99$^{th}$ percentiles and defined in Table A1. *p-values* computed using standard errors clustered at the bank level are reported in parentheses.

| | ΔZ_SCORE | ΔMES_VAR5% | ΔSRISK |
|---|---|---|---|
| | (1) | (2) | (3) |
| CYBERLAW | 0.059 | -0.029 | -0.028 |
| | [0.027] | [0.004] | [0.082] |
| ΔBANK_SIZE | -0.016 | 0.015 | -0.010 |
| | [0.687] | [0.633] | [0.765] |
| ΔEQUITY_RATIO | -1.999 | -0.030 | 0.520 |
| | [0.000] | [0.879] | [0.104] |
| ΔDEPOSIT_RATIO | 0.021 | -0.147 | 0.141 |
| | [0.845] | [0.178] | [0.205] |
| ΔCOST_EFFICIENCY | 0.110 | 0.027 | 0.074 |
| | [0.010] | [0.214] | [0.074] |
| ΔREV_DIV | 0.033 | -0.009 | -0.22 |
| | [0.636] | [0.807] | [0.053] |
| ΔEXP_DEP_INSUR | 0.094 | -0.005 | 0.006 |
| | [0.131] | [0.892] | [0.845] |
| ΔRULE OF LAW | 0.013 | 0.001 | -0.029 |
| | [0.793] | [0.999] | [0.348] |
| ΔLOG_GDP | 0.188 | 0.101 | 0.001 |
| | [0.088] | [0.001] | [0.969] |
| ΔLOG_GDPPERCAP | -0.367 | -0.086 | -0.007 |
| | [0.018] | [0.035] | [0.888] |
| ΔCREDIT_TO_GDP | 0.097 | -0.030 | 0.051 |
| | [0.014] | [0.220] | [0.455] |
| | | | |
| Observations | 71,449 | 3,158 | 4,126 |
| Adjusted R-squared | 0.04 | 0.03 | 0.02 |
| Country FE, Year FE | Yes | Yes | Yes |
| Cluster by Bank | Yes | Yes | Yes |

Table IA14: Change Analysis

This table provides regression results of changes in banking system stability, measured by the change in the Z-index (ΔZSCORE), marginal expected shortfall calculated at 5th percentile of value at risk (ΔMES_VAR5%), and systemic risk (ΔSRISK) on cybercrime law enactment (CYBERLAW) in Columns (1)-(3), respectively. In terms of control variables, they consist of changes in bank size (ΔBANK_SIZE), the equity-to-total-assets ratio (ΔEQUITY_RATIO), the deposit-to-total-assets ratio (ΔDEPOSIT_RATIO), the cost-to-income ratio (ΔCOST_EFFICIENCY), revenue diversification (ΔREV_DIV), explicit deposit insurance (ΔEXP_DEP_INSUR), rule of law (ΔRULE OF LAW), log GDP (ΔLOG_GDP), log GDP per capita (ΔLOG_GDPPERCAP), and the credit-to-GDP ratio (ΔCREDIT_TO_GDP). All continuous variables are winsorized at the 1st and 99th percentiles and defined in Table A1. *p-values* computed using standard errors clustered at the bank level are reported in parentheses.

| | Table IA15: Non-US Sample | | |
|---|---|---|---|
| | Z_SCORE | MES_VAR5% | SRISK |
| | (1) | (2) | (3) |
| CYBERLAW | 0.097 | -0.006 | -0.036 |
| | [0.000] | [0.004] | [0.025] |
| Control Variables | Yes | Yes | Yes |
| Observations | 80,639 | 6,307 | 5,925 |
| Adjusted R-squared | 0.231 | 0.369 | 0.334 |
| Country FE, Year FE | Yes | Yes | Yes |
| Cluster by Bank | Yes | Yes | Yes |

This table provides regression results of different banking system stability measures (i.e., *Z_SCORE, MES_VAR5%,* and *SRISK*) on cybercrime law enactment (*CYBERLAW*) in Columns (1)-(3), respectively, without U.S. banks. In terms of control variables, they consist of bank size (BANK_SIZE), the equity-to-total-assets ratio *(EQUITY_RATIO*), the deposit-to-total-assets ratio (*DEPOSIT_RATIO*), the cost-to-income ratio (*COST_EFFICIENCY*), revenue diversification (*REV_DIV),* rule of law (*RULE OF LAW),* log GDP *(LOG_GDP),* log GDP per capita *(LOG_GDPPERCAP),* and the credit-to-GDP ratio *(CREDIT_TO_GDP)*. All continuous variables are winsorized at the 1$^{st}$ and 99$^{th}$ percentiles and defined in Table A1. *p-values* computed using standard errors clustered at the bank level are reported in parentheses.

| | Table IA16: Non-GFC Period | | |
|---|:---:|:---:|:---:|
| | Z_SCORE | MES_VAR5% | SRISK |
| | (1) | (2) | (3) |
| CYBERLAW | 0.015 | -0.007 | -0.028 |
| | [0.015] | [0.063] | [0.070] |
| Control Variables | Yes | Yes | Yes |
| Observations | 152,675 | 5,451 | 7,261 |
| Adjusted R-squared | 0.160 | 0.362 | 0.313 |
| Country FE, Year FE | Yes | Yes | Yes |
| Cluster by Bank | Yes | Yes | Yes |

This table provides regression results of different banking system stability measures (i.e., *Z_SCORE, MES_VAR5%,* and *SRISK*) on cybercrime law enactment (*CYBERLAW*) in Columns (1)-(3), respectively, without the global financial crisis periods (2007-2008). In terms of control variables, they consist of bank size (*BANK_SIZE*), the equity-to-total-assets ratio *(EQUITY_RATIO)*, the deposit-to-total-assets ratio (*DEPOSIT_RATIO*), the cost-to-income ratio (*COST_EFFICIENCY*), revenue diversification (*REV_DIV),* rule of law (*RULE OF LAW),* log GDP *(LOG_GDP),* log GDP per capita *(LOG_GDPPERCAP)*, and the credit-to-GDP ratio *(CREDIT_TO_GDP)*. All continuous variables are winsorized at the 1$^{st}$ and 99$^{th}$ percentiles and defined in Table A1. *p-values* computed using standard errors clustered at the bank level are reported in parentheses.

| | Table IA17: Additional Control Variables | | |
|---|---|---|---|
| | Z_SCORE | MES_VAR5% | SRISK |
| | (1) | (2) | (3) |
| CYBERLAW | 0.030 | -0.015 | -0.037 |
| | [0.012] | [0.054] | [0.048] |
| BANK_SIZE | 0.002 | 0.030 | 0.049 |
| | [0.687] | [0.000] | [0.001] |
| EQUITY_RATIO | 1.345 | 0.143 | -0.124 |
| | [0.000] | [0.000] | [0.345] |
| DEPOSIT_RATIO | 0.552 | -0.002 | -0.289 |
| | [0.000] | [0.849] | [0.003] |
| COST_EFFICIENCY | -0.832 | 0.041 | 0.145 |
| | [0.000] | [0.000] | [0.032] |
| REV_DIV | -0.951 | -0.008 | -0.223 |
| | [0.000] | [0.497] | [0.024] |
| EXP_DEP_INSUR | 0.104 | 0.007 | 0.025 |
| | [0.001] | [0.456] | [0.603] |
| RULE OF LAW | 0.075 | -0.002 | 0.048 |
| | [0.066] | [0.917] | [0.251] |
| LOG_GDP | -0.728 | -0.144 | -0.04 |
| | [0.000] | [0.000] | [0.760] |
| LOG_GDPPERCAP | 0.816 | -0.079 | -0.008 |
| | [0.000] | [0.028] | [0.949] |
| CREDIT_TO_GDP | -0.069 | 0.044 | 0.099 |
| | [0.070] | [0.046] | [0.095] |
| ACTRES | -0.008 | -0.006 | -0.007 |
| | [0.056] | [0.001] | [0.017] |
| CAPREG | 0.029 | 0.010 | 0.022 |
| | [0.012] | [0.230] | [0.037] |
| SUPOWER | -0.003 | 0.001 | 0.002 |
| | [0.505] | [0.738] | [0.735] |
| | | | |
| Observations | 161,698 | 5,608 | 7,316 |
| Adjusted R-squared | 0.153 | 0.382 | 0.307 |
| Country FE, Year FE | Yes | Yes | Yes |
| Cluster by Bank | Yes | Yes | Yes |

This table provides regression results of different banking system stability measures (i.e., *Z_SCORE, MES_VAR5%,* and *SRISK*) on cybercrime law enactment (*CYBERLAW*) in Columns (1)-(3), respectively. In terms of control variables, they consist of bank size (BANK_SIZE), the equity-to-total-assets ratio *(EQUITY_RATIO)*, the deposit-to-total-assets ratio *(DEPOSIT_RATIO)*, the cost-to-income ratio (*COST_EFFICIENCY*), revenue diversification (*REV_DIV),* rule of law (*RULE OF LAW),* log GDP *(LOG_GDP),* log GDP per capita *(LOG_GDPPERCAP)*, and the credit-to-GDP ratio *(CREDIT_TO_GDP)*. Additional control variables include bank activity restrictions (*ACTRES*), bank capital regulation (*CAPREG*), and bank supervisory power (*SUPOWER*). All continuous variables are winsorized at the 1st and 99th percentiles and defined in Table A1. *p-values* computed using standard errors clustered at the bank level are reported in parentheses.

| | Table IA18: Additional DiD Analysis | | |
|---|---|---|---|
| | Z_SCORE | MES_VAR5% | SRISK |
| | (1) | (2) | (3) |
| POST*TREATMENT | 0.048 | -0.047 | -0.016 |
| | [0.022] | [0.006] | [0.045] |
| POST | 0.034 | -0.002 | -0.003 |
| | [0.174] | [0.815] | [0.789] |
| TREATMENT | -0.161 | 0.055 | 0.011 |
| | [0.014] | [0.298] | [0.022] |
| Control Variables | Yes | Yes | Yes |
| Observations | 30,009 | 1,690 | 1,507 |
| Adjusted R-squared | 0.321 | 0.406 | 0.564 |
| Cohort*Country FE | Yes | Yes | Yes |
| Cohort*Year FE | Yes | Yes | Yes |
| Cluster by Bank | Yes | Yes | Yes |

This table reports the results of the staggered difference-in-differences regressions using stacked event-by-event models suggested by Baker et al., (2022) after we remove countries enacting significant bank regulation reforms within a five-year window of the cybercrime law enactment. TREATMENT is an indicator variable, which equals 1 for countries enacting cybercrime law and zero otherwise during the sample periods. POST is a dummy variable equal to 1 if one-to-five years post law enactment and zero otherwise. Columns (1)-(4) report the results associated with the three measures of banking system stability, including the Z-index (Z_SCORE), marginal expected shortfall calculated at the 5$^{th}$ percentile of value at risk (MES_VAR5%), and systemic risk (SRISK). In terms of control variables, they consist of bank size (BANK_SIZE), the equity-to-total-assets ratio (EQUITY_RATIO), the deposit-to-total-assets ratio (DEPOSIT_RATIO), the cost-to-income ratio (COST_EFFICIENCY), revenue diversification (REV_DIV), rule of law (RULE OF LAW), log GDP (LOG_GDP), log GDP per capita (LOG_GDPPERCAP), and the credit-to-GDP ratio (CREDIT_TO_GDP). p-values computed using standard errors clustered at the bank level are reported in parentheses.

|  | Table IA19: Additional Subsample Analysis | | |
|---|---|---|---|
|  | *Z_SCORE* | *MES_VAR5%* | *SRISK* |
|  | (1) | (2) | (3) |
| *CYBERLAW* | 0.019 | -0.021 | -0.034 |
|  | [0.036] | [0.004] | [0.036] |
| Control Variables | Yes | Yes | Yes |
| Observations | 174,573 | 5,890 | 8,094 |
| Adjusted R-squared | 0.149 | 0.380 | 0.300 |
| Country FE | Yes | Yes | Yes |
| Year FE | Yes | Yes | Yes |
| Cluster by Bank | Yes | Yes | Yes |

This table provides regression results of different banking system stability measures (i.e., *Z_SCORE, MES_VAR5%,* and *SRISK*) on cybercrime law enactment (*CYBERLAW*) in Columns (1)-(3), respectively. In this sample, we exclude countries that enacted cybercrime laws in direct response to major cyber incidents (e.g., Afghanistan, Brazil, Cameroon, Czech Republic, Nigeria, Saudi Arabia, Singapore, Slovenia, Sudan, Ukraine, Vietnam). In terms of control variables, they consist of bank size (*BANK_SIZE*), the equity-to-total-assets ratio (*EQUITY_RATIO*), the deposit-to-total-assets ratio (*DEPOSIT_RATIO*), the cost-to-income ratio (*COST_EFFICIENCY*), revenue diversification (*REV_DIV),* rule of law (*RULE OF LAW),* log GDP *(LOG_GDP),* log GDP per capita *(LOG_GDPPERCAP),* and the credit-to-GDP ratio *(CREDIT_TO_GDP)*. All continuous variables are winsorized at the 1$^{st}$ and 99$^{th}$ percentiles and defined in Table A1. *p-values* computed using standard errors clustered at the bank level are reported in parentheses.

**Figure IA1: Developed vs Developing Countries**

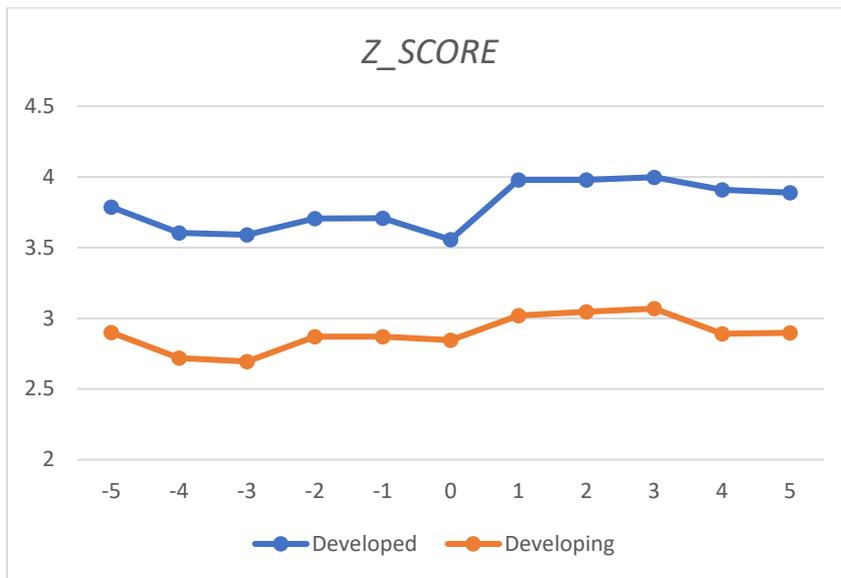

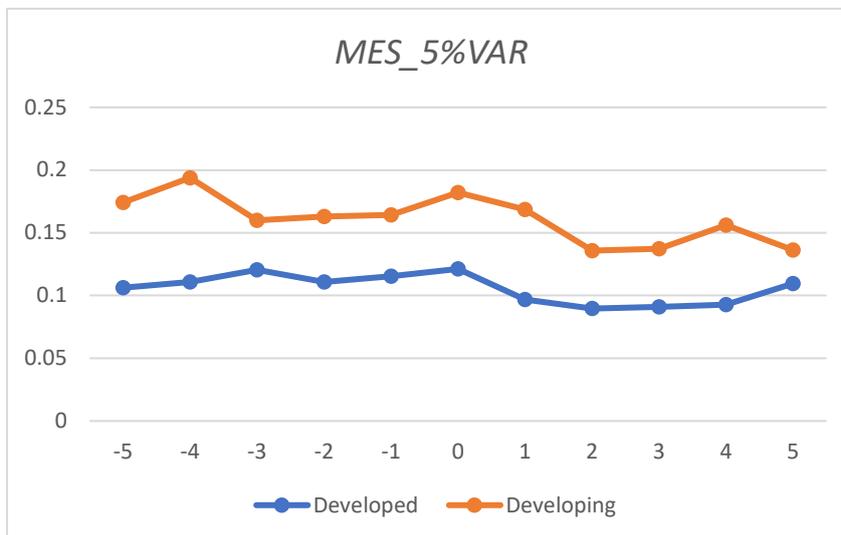

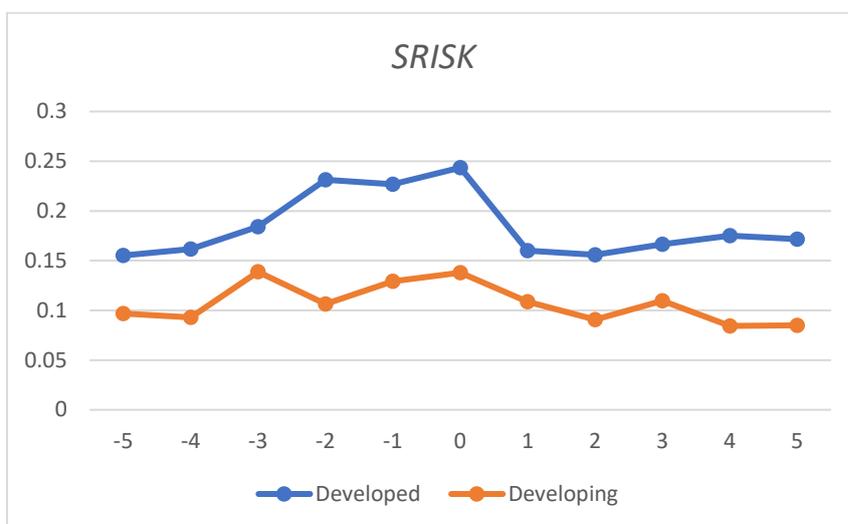

This figure shows the variations across different measures of bank stability surrounding the enactment of cybercrime laws in developed and developing countries.

# Figure IA2: Bootstrapping Sampling Distributions

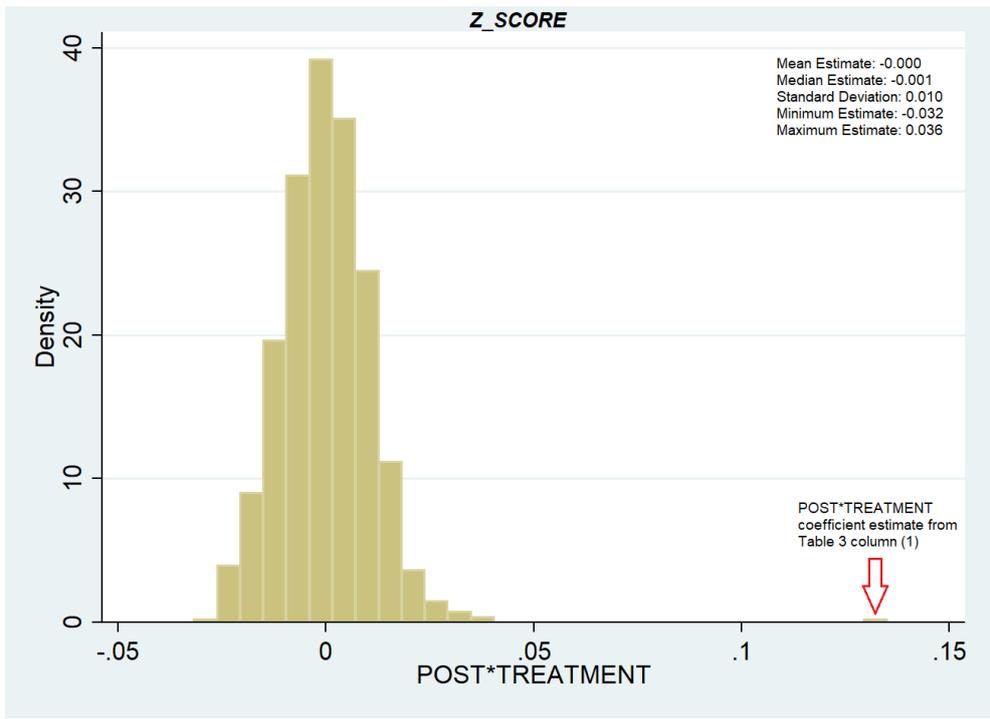

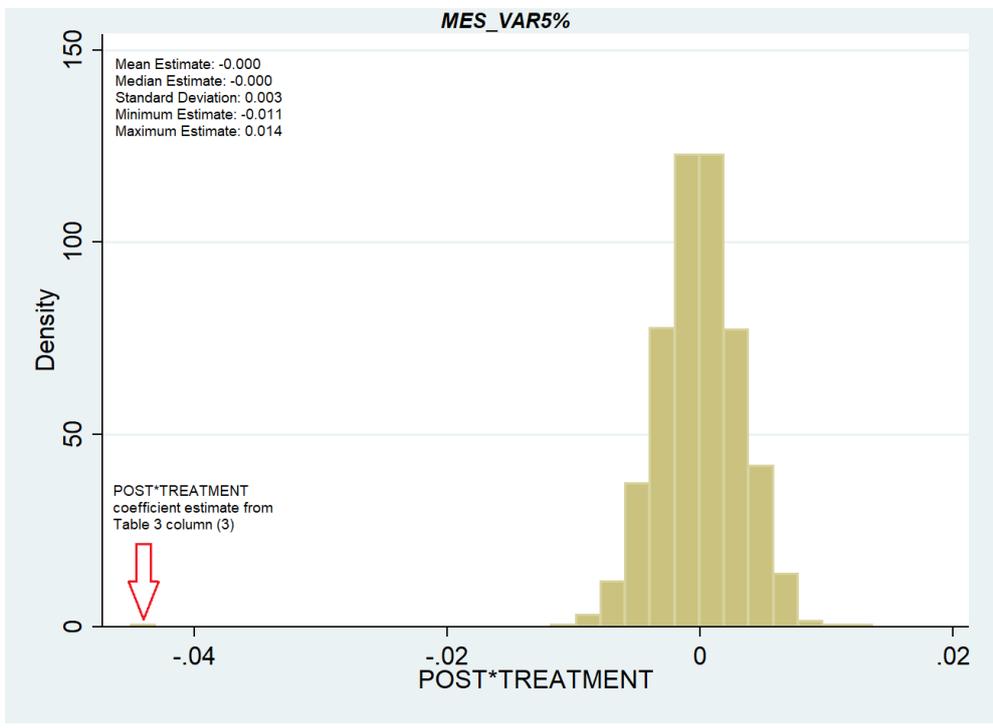

**Figure IA2: Bootstrapping Sampling Distributions (Cont.)**

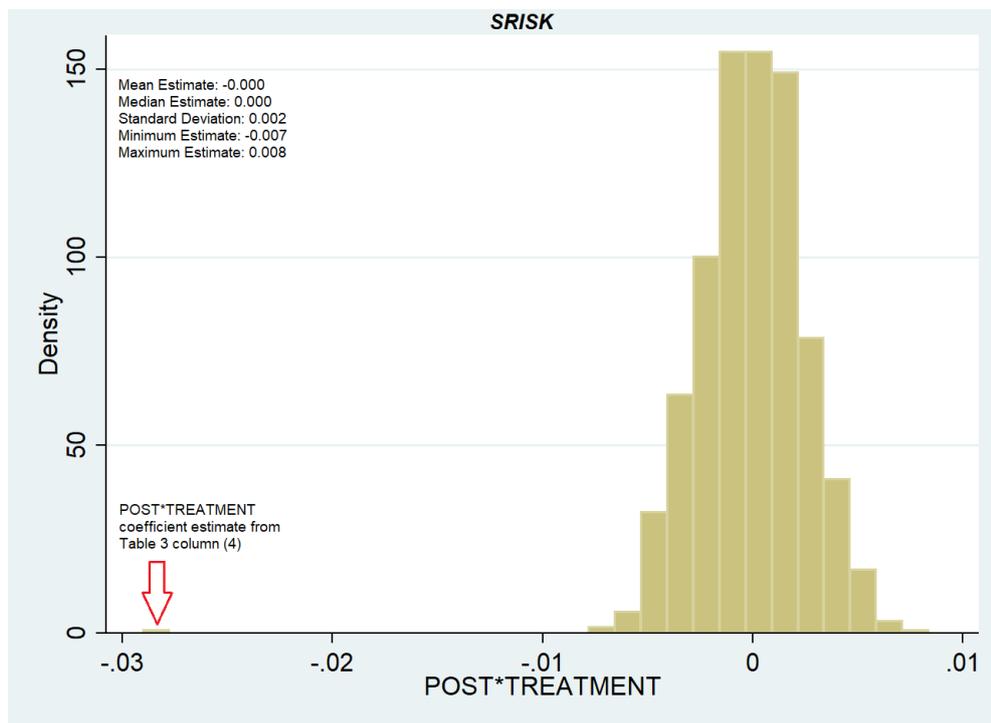

This figure plots the bootstrapping sampling distributions of coefficients on three measures of bank stability obtained from 1,000 simulations.